\documentclass[twocolumn]{aastex631}
\usepackage{color,comment}
\usepackage{enumitem}
\usepackage[T1]{fontenc}
\usepackage{tabularray}
\hypersetup{linkcolor=blue,citecolor=blue,urlcolor=blue}

\newcommand{\secref}[1]{\S\ref{#1}}
\graphicspath{{./}{figures/}}
\usepackage{float}
\usepackage{amsmath}
\usepackage{multirow}
\usepackage{rotating}
\usepackage{bm}
\shortauthors{Garg et al.}

\begin{document}

\title{Theoretical strong line metallicity diagnostics for the  JWST era}

\correspondingauthor{Prerak Garg}
\email{prerakgarg@ufl.edu}


\author[0000-0002-5923-2151]{Prerak Garg}
\affil{Department of Astronomy, University of Florida, 211 Bryant Space Sciences Center, Gainesville, FL 32611, USA}

\author[0000-0002-7064-4309]{Desika Narayanan}
\affil{Department of Astronomy, University of Florida, 211 Bryant Space Sciences Center, Gainesville, FL 32611, USA}
\affil{University of Florida Informatics Institute, 432 Newell Drive, CISE Bldg E251, Gainesville, FL 32611, USA}
\affil{Cosmic Dawn Center at the Niels Bohr Institute, University of Copenhagen and DTU-Space, Technical University of Denmark}

\author[0000-0003-4792-9119]{Ryan L. Sanders}
\affil{Department of Physics and Astronomy, University of California, Davis, One Shields Ave, Davis, CA 95616, USA}
\affil{Department of Physics and Astronomy, University of Kentucky, 505 Rose Street, Lexington, KY 40506, USA}

\author[0000-0003-2842-9434]{Romeel Dav\'e}
\affil{Institute for Astronomy, Royal Observatory Edinburgh, EH9 3HJ, UK}
\affil{University of the Western Cape, Bellville, Cape Town 7535, South Africa}
\affil{South African Astronomical Observatory, Cape Town 7925, South Africa}

\author[0000-0003-1151-4659]{Gerg\"o Popping}
\affil{European Southern Observatory, Karl-Schwarzschild-Strasse 2, D-85748, Garching, Germany}

\author[0000-0003-3509-4855]{Alice E. Shapley}
\affil{Department of Physics and Astronomy, University of California, Los Angeles, 430 Portola Plaza, Los Angeles, CA 90095, USA}

\author[0000-0001-6106-5172]{Daniel P. Stark}
\affil{Steward Observatory, University of Arizona, 933 N Cherry Ave, Tucson, AZ 85721, USA}

\author[0000-0002-1410-0470]{Jonathan R. Trump}
\affil{Department of Physics, 196 Auditorium Road, Unit 3046, University of Connecticut, Storrs, CT 06269 USA}

\begin{abstract}
The ratios of strong rest-frame optical emission lines are the dominant indicator of metallicities in high-redshift galaxies. Since typical strong-line based metallicity indicators are calibrated on auroral lines at $z=0$, their applicability for galaxies in the distant Universe is unclear. In this paper, we make use of mock emission line data from cosmological simulations to investigate the calibration of rest-frame optical emission lines as metallicity indicators at high redshift. Our model, which couples the {\sc simba} cosmological galaxy formation simulation with \textsc{cloudy} photoionization calculations, includes contributions from H\scalebox{0.9}{ II} regions, post-AGB stars and Diffuse Ionized Gas (DIG). We find mild redshift evolution in the 12 indicators that we study, which implies that the dominant physical properties that evolve in our simulations do have a discernible impact on the metallicity calibrations at high redshifts. When comparing our calibrations with high redshift auroral line observations from James Webb Space Telescope we find a slight offset between our model results and the observations and find that a higher ionization parameter at high redshifts can be one of the possible explanations. We explore the physics that drives the shapes of strong-line metallicity relationships and propose calibrations for hitherto unexplored low-metallicity regimes. Finally, we study the contribution of DIG to total line fluxes. We find that the contribution of DIG increases with metallicity at z $\sim$ 0 for singly ionized oxygen and sulfur lines and can be as high as 70\% making it crucial to include their contribution when modeling nebular emission. 
\end{abstract}


\section{Introduction} 
Estimating gas-phase metallicity is crucial in understanding the chemical evolution of galaxies across cosmic time. The metal content of galaxies is regulated by enrichment through stellar evolution and the inflow and outflow of gas to the intergalactic medium. Historically, nebular emission lines have been used as one of the primary methods for estimating galaxy metallicities. These lines, first discovered in 1864 by Sir William Huggins, in an early spectrum of the planetary nebula, NGC 6543 (the Cat's Eye Nebula). In the spectrum, Huggins saw a set of strong green nebular emission lines at 4959 and 5007 {\AA}. Unable to deduce their source he proposed that they must be originating from an unknown element, which was termed nebulium. Over a half-century later, \citet{bowen1927} found that these lines instead arise from the transition of doubly ionized oxygen (O++), and no new element was needed to explain them. It is now understood that many strong nebular lines primarily arise from the forbidden transition of ionized metal species.

Over the last century, significant effort has been expended to employ these nebular emission lines as diagnostics of the physical conditions in galaxies. One such important tool is the use of rest-frame optical nebular emission line ratios to derive ionized gas metallicity. The most accurate method of deriving metallicity in external galaxies is the so-called "direct" method, which uses temperature-sensitive auroral lines like [S\scalebox{0.9}{ II}]{$\lambda$4072}, [O\scalebox{0.9}{ III}]{$\lambda$4363}, [N\scalebox{0.9}{ II}]{$\lambda$5755}, and [S\scalebox{0.9}{ III}]{$\lambda$6312} to derive electron temperature $(T_e)$\citep[e.g.][]{McLennan1925, Keenan1996} which when combined with an ionization correction factor can then be used to derive an accurate measurement of ionized gas metallicity \citep{Pagel1992, osterbrook2006, Izotov2006, PrezMontero_2017}. Recombination-based methods are even better, but they prove too challenging for the majority of extra-galactic sources. Numerous studies have used the "direct" method to derive ionized gas phase metallicity in galaxies \citep{Bresolin_2007, Monreal2012, Westmoquette2013, andrews2013, Maseda_2014, Berg_2015, McLeod2015, Croxall_2015, Croxall_2016, Lin_2017}. Unfortunately, these lines are relatively weak, which makes them extremely hard to observe in all but the nearest galaxies; indeed, auroral line measurements have only recently been taken in about 60 galaxies at redshifts $z \gtrapprox 1$ \citep{villar2004, Yuan_2009, erb2010, rigby2011, Brammer_2012, Christensen2012a, Christensen2012b, Stark2014, Bayliss_2014, James2014, Jones2015, steidel2014, Sanders_2016, Kojima2017, Berg_2018, Patricio2018, Sanders_direct2019, Gburek_2019, curti_2022, trump_2022, nakajima_2023, sanders_direct_2023}. The fact that almost half of these come from JWST, in the last year shows that it is still non-trivial to get large and representative auroral line measurements at $z \gtrapprox 1$ because current samples are still biased toward a high specific star formation rate (sSFR) relative to the galaxy main sequence.

To overcome this drawback, several studies have calibrated auroral line-based metallicity measurements against ratios of bright rest-frame optical emission line involving,[N\scalebox{0.9}{ II}]{$\lambda 6583$}, [O\scalebox{0.9}{ III}]$\lambda 5007$, [O\scalebox{0.9}{ II}]$\lambda 3727, 29$, and [S\scalebox{0.9}{ II}]$\lambda 6717, 31$, Ne\scalebox{0.9}{ III}]$\lambda 3869$, H$\alpha$, and H$\beta$   (more commonly called "strong lines") \citep[e.g.,][]{mccall1985, mcgaugh1991, zaritsky1994, kewley2002, Pettini, Pilyugin2005, Maiolino2008, dopita2016, curti_2017, curti_2020}. These strong line calibrations are now widely used for deriving metallicity measurements in galaxies where it is hard to observe the auroral lines. However, since these were calibrated using local auroral lines observations, their applicability at high redshift remains in question as the ISM ionization conditions might evolve with redshift \citep{steidel2016,strom2018,Shapley_2019,Topping2020a, Sanders_direct2019}.

Tackling this issue has now taken center stage with the launch of the James Webb Space Telescope (JWST). Already in its first year, a number of studies  \citep[e.g.][]{arellano-cordova_2022, brinchmann_2022, curti_2022, katz_2022, langeroodi_2022, matthee_2022, rhoads_2023, schaerer_2022, sun_2022, tacchella_2022, taylor_2022, trump_2022, wang_2022, heintz_2023, nakajima_2023, sanders_2023, sanders_direct_2023, tang_2023, williams_2023} have observed nebular emission lines from high redshift galaxies using JWST/NIRSpec. The goal of this paper is to assess whether locally-calibrated strong line metallicity estimators apply in the physical conditions appropriate for high-$z$ galaxies. The paper is divided as follows: In \secref{sec:model_description} we describe our methodology. In \secref{sec:local_mi} we compare our model calibrations against locally observed auroral line based calibrations from the literature. In \secref{sec:z_mi} we look at the redshift evolution of strong line metallicity calibrations, propose a new metallicity calibration and look at the relative contribution of different ionizing sources. In \secref{sec:discussion} we compare our results in context to other relevant studies and go over some of our model caveats. In \secref{sec:summary} we summarize our main conclusions.Throughout, we assume a $\Lambda$ cold dark matter cosmology with $\Omega_m = 0.3$, $\Omega_{\Lambda} = 0.7$, and $H_0 = 68$ kms$^{-1}$ Mpc$^{-1}$  

\vspace{5mm}
\section{Model Description}\label{sec:model_description}
\subsection{Model Overview}
To accurately estimate nebular emission in a galaxy we need to precisely model a variety of different physical processes. For all the photoionization calculations we make use of \textsc{cloudy} version 17.00 \citep{ferland2017}. We model three main sources of ionizing radiation within the context of galaxy evolution simulations:  H\scalebox{0.9}{ II} regions, post-AGB stars, and Diffuse Ionized Gas (DIG). The H\scalebox{0.9}{ II} regions and post-AGB stars are modeled as constant-density spherical shells with a fixed escape fraction. The escaping ionizing photons then ionize the surrounding gas particles which give rise DIG, modeled as gas cells with a plane-parallel geometry. In the following subsections, we describe in detail our modeling strategy for each of these sources, our assumptions, and the model parameters we use for each of them.

\subsection{\textsc{simba} simulations}
We first simulate galaxies using the \textsc{simba} hydrodynamic cosmological galaxy formation simulation \citep{simba}. \textsc{simba} is an updated version of \textsc{mufasa} hydrodynamic simulation \citep{dave2016mufasa}, and uses the \textsc{gizmo} hydrodynamic solver in the meshless finite mass mode. {\sc simba} uses the \textsc{grackle-3.1} library \citep{smith2017grackle} to calculate photoionization and non-equilibrium radiative cooling from H, He, and metals. A self-shielding prescription from \citet{rahmati_2013} is applied to a uniform ionizing background \citep{haardt_2012}. Star formation occurs in H$_2$ gas, employing a volumetric Schmidt-Kennicutt relation \citep{kennicutt_1998} which is calculated using the \citet{krumholz2009} sub-resolution model for computing the H$_2$ gas fraction with some modifications (see \citet{dave2016mufasa}). The stellar feedback is modeled as a two-phased decoupled kinetic outflow with 30\% hot component \citep{dave2016mufasa}. The mass loading factors in stellar winds are taken from calibrations from the FIRE simulations as measured by \citet{angles2017b}. Black holes are seeded in galaxies with stellar mass $M_* > 10^{9.5} M_{\odot}$. Black hole growth is driven via two accretion modes: Bondi accretion \citep{bondi1944}, only from the hot halo, and torque-driven cold accretion \citep{angles2017a}. \textsc{simba} can accurately reproduce many of the observed galaxy scaling relations including the star-forming main sequence, galaxy stellar mass function, gas-phase, and stellar mass-metallicity relation across redshifts. 

One important feature that differentiates \textsc{simba} from other cosmological simulations is that the model includes an on-the-fly dust production and destruction model. Dust is produced via Type II supernovas and AGB stars, whereas dust can either be destroyed via thermal sputtering and shocks or can be consumed during star formation. \citet{li2019} demonstrated that this model can successfully match the observed dust-to-gas and dust-to-metals ratios at low and high-$z$, as well as the cosmic evolution of $\Omega_{\rm dust}$ \citep{peroux_cosmic_2020}. \citet{lovell_2021} showed that it can also match the sub-millimeter galaxy abundances pretty well. Having a sophisticated model of dust production and destruction is important in accurately modeling nebular emission since some metals are locked up in dust and are therefore not available for ionization.

We run a total of 4 \textsc{simba} boxes each with a different random seed till $z \sim 0$ with a simulation box size of side length $25 h^{-1}$ Mpc with $512^3$ particles giving us a baryon resolution of $2.3\times10^6$ M$_{\odot}$. The snapshots from \textsc{simba} are first processed through \textsc{caesar} \footnote{\url{https://github.com/dnarayanan/caesar}} galaxy catalog generator \citep{Thompson2014}. Once we have identified all the galaxies in a snapshot, we filter out those that do not have at least one star younger than 10 Myr. Enforcing this criterion ensures that our sample galaxies always have some young stars, the major contributor to the ionizing flux in our model. With star particle mass of $2.3\times10^6$ M$_{\odot}$, this means that a galaxy has to have an average star formation rate of 0.23 M$_{\odot}$/yr to have a reasonable probability of having one such star particle. Many small (but still resolved) main sequence galaxies fall below this. Thus, due to the stochastic nature of the star formation algorithm in \textsc{simba} will exclude a lot of small galaxies, which actually should have emission lines. We find that the potentially star forming galaxies that we exclude from our sample due this cutoff is about 5 - 10\% of at M$_*$ > 9.5 increasing to around 60 - 80\% at  M$_*$ < 8.0. 

\subsection{Photoionization modeling}
The nebular emission model that we use here is an updated one to that presented in \citet{garg2022bpt}, with a few notable changes. In particular, while \citet{garg2022bpt} only considered nebular emission from H\scalebox{0.9}{ II} regions around young stars, we have now expanded the model to also include contributions of ionized gas around post-AGB stars DIG. In the following subsections, we describe in detail our modeling parameters and how we model each of the three sources of nebular emission (H\scalebox{0.9}{ II} regions, DIG, and the regions surrounding post-AGB stars). 

\subsubsection{Stellar Model parameters}
The mass resolution of our parent simulations is of the order of $10^6$ M$_{\odot}$. Because this is larger than the typical mass of observed clusters \citep{chandar2014, chandar2016, linden2017, larson2020}, we employ a sub-resolution model for breaking these star particles into smaller sub-units:

\paragraph{Cluster Mass Distribution}
If we were to assume a single a H\scalebox{0.9}{ II} region around our massive $2.3\times10^6$ M$_\odot$ clusters, this would result in an overestimation of the ionizing flux per unit H\scalebox{0.9}{ II} region. We, instead, consider each star particle in our simulation to represent a collection of unresolved clusters with individual H\scalebox{0.9}{ II} regions. We subdivide each star particle into a collection of smaller mass star particles with the same age and metallicity as the parent star particles but their masses are derived from a cluster mass distribution of the form

\begin{equation}\label{eq:cmdf_eqn}
\frac{dN}{dM} \propto M^{\beta}
\end{equation}
Where $M$ is the stellar mass of the star particle, $N$ is the number of star particles, and $\beta$ is the power-law index. While observations have indicated power law-like mass distributions for observed clusters, there is yet no consensus as to the powerlaw index, $\beta$. Observations suggest a range from $-2.2<\beta<-1.7$ \citep{chandar2014, chandar2016, linden2017, larson2020}. We, therefore, assume $\beta=-2.0$. We subdivide each star particle into 4 bins with a mass range from $10^{3.5}-10^{5.0}$ M$_\odot$. The distribution is done in such a way that the total mass of all the sub-particles sums to the mass of the parent star particle ensuring that the total mass is conserved.

\paragraph{Age Distribution}
Since clusters within the parent particle that was subdivided may have a distribution of ages, we further subdivide each of the smaller mass star clusters into multiple individual particles with varying ages based on a power law distribution function of the form:
\begin{equation}\label{eq:age_eqn}
\frac{dN}{d\tau} \propto \tau^{\gamma}
\end{equation}
Where $\tau$ is the age of the star particle, $N$ is the number of star particles and $\gamma$ is the power-law index. Observations by \citet{chandar2014,chandar2016} suggest that $\gamma$ can vary between -0.55 to -0.74 with only a modest dependence on cluster mass. We, therefore, fix $\gamma=-0.65$. We divide all the star particles with ages between 2 to 10 Myr into 4 bins. The age distribution is constructed to ensure that the mean age of all the sub-particles is as close age of the parent particle as possible. We apply a threshold of 0.1 Myr and if the difference between the average age and the age of the parent star particle is above the threshold then the width (difference between the start and end age of the distribution) is decreased at each step until the threshold is met. This is done at most 100 times and if we fail to meet the threshold even after that then the particle is not broken down.

\begin{figure}[htp]
\centering
  \includegraphics[width=\columnwidth]{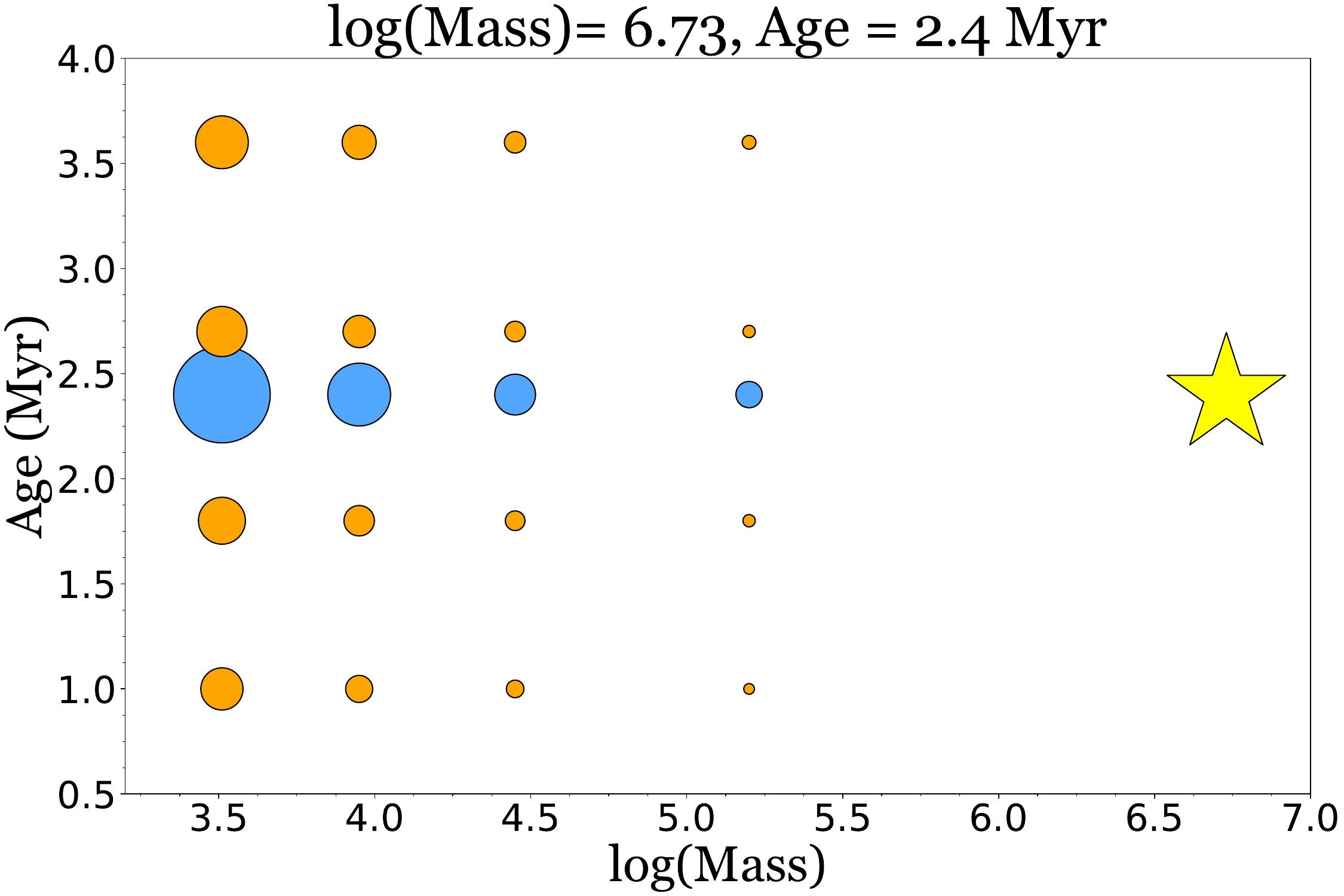}
  \caption{An example of our methodology for dividing star particles into a collection of sub-particles with varying age and mass based on power law distributions. The parent star particle (yellow star) has a mass of $10^{6.73}$ $M_\odot$ and an age of 2.4 Myr. The particle is first subdivided into 4 mass bins (blue circles) and each of these sub-particles is then further subdivided into 4 age bins (orange circles). The relative area of the circle represents how many stars are there in each bin. In the end, this one-star particle is broken down into 400-star particles spread across 16 bins of varying ages and masses.}
  \label{fig:age_dist}
\end{figure}

As an example, in Figure \ref{fig:age_dist} we show how a single star particle of mass of $10^{6.73}$ $M_\odot$ and age of 2.4 Myr is subdivided into multiple star particles with varying mass and age based on the power law distributions described above. The parent star particle (yellow star) is first broken down into 4 mass bins, shown by the blue circles with their relative area representing how many particles are there in each bin. The particles in each of those mass bins (blue circles) are then further subdivided into 4 age bins based on the age distribution function, shown by the orange circles. Thus, in the end, one-star particle gets broken down into $\sim 400$ particles spread over $16$ age and mass bins.  

\paragraph{Spectral library and isochrones}
For all the analysis in this paper unless otherwise specified we make use of binary models grids from Binary Population and Spectral Synthesis (\textsc{bpass}) library \citet{eldridge2017} and \citet{stanway2018} with Chabrier initial mass function (IMF) and a 100 $M_\odot$ cutoff. 

\subsubsection{Sources of ionizing radiation}\label{sec:method_sources}
\paragraph{H\scalebox{0.9}{ II} regions}
H\scalebox{0.9}{ II} regions around young stars are the primary source of ionizing radiation in a star-forming galaxy. To model H\scalebox{0.9}{ II} regions we first take all the star particles younger than 10 Myr and subdivide them based on the cluster mass and age distribution functions as described above. We apply this age cutoff to ensure that only stars that produce an adequate amount of ionizing flux are considered in our calculation (for computational efficiency). 

We model H\scalebox{0.9}{ II} regions as spherical shells around all the individual sub-particles of interest and compute the emergent nebular line emission from these H\scalebox{0.9}{ II} regions using {\sc cloudy} \citep{ferland2017}. The model inputs for the {\sc cloudy} photoionization calculations are the shape of the ionizing radiation field, the ionizing source luminosity, gas phase abundance, hydrogen density, and the inner radius. To compute the ionizing radiation field, the SEDs for all subdivided star particles are computed using Flexible Stellar Population Synthesis (\textsc{fsps})\footnote{\url{https://github.com/cconroy20/fsps}} \citep{conroy2009,conroy2010}. We model each star particle as a simple stellar population (SSP) and set the stellar metallicity to log(Fe/H) rather than the total metal content by mass so as to mimic the effects of $\alpha$-enhancement \citep{garg2022bpt}. The age of the star is set based on the age returned from the cosmological simulation.

The ionizing flux for each star particle is computed via:
\begin{equation}\label{q_eqn}
Q = M_*\int_{\nu{_0}}^{\infty}{\frac{L_\nu}{h\nu} d_\nu}    
\end{equation}
Where $L_\nu$ is in units of ergs/s/Hz and is taken directly from the \textsc{fsps} SED, $M_*$ is the stellar mass, and Q is the rate of ionizing photons in units of $s^{-1}$. The gas phase abundances are assumed to be the same as that of the star particle and are taken directly from the simulation. \textsc{simba} keeps track of the abundances of 10 elements (He, C, N, O, Ne, Mg, Si, S, Ca, and Fe) apart from hydrogen, and for all the elements except nitrogen, we make use of the abundances tracked by the simulation. Nitrogen production is fairly complicated to model since it can be produced via primary or secondary nucleosynthesis and \textsc{simba} does a fairly poor job of matching the observed (N/O) vs (O/H) relation from \citet{pilyugin2012} (Equation \ref{pilyugin_eqn}, see \citet{garg2022bpt} and \cite{hough_simba-c_2023} for more info). Thus, we set nitrogen abundance in our model according to Equation \ref{pilyugin_eqn}. 

\begin{equation}\label{pilyugin_eqn}
\begin{aligned}
\mathrm{log (N/O)} &= -1.493 \\
&\text{ for } 12 + \mathrm{log(O/H)} < 8.14, \\
&= 1.489 \times [12 + \mathrm{log(O/H)}] - 13.613\\
&\text{ for } 12 + \mathrm{log(O/H)} > 8.14
\end{aligned}
\end{equation}

We assume fixed physical conditions for all H\scalebox{0.9}{ II} regions in our simulations, irrespective of redshift. This includes an assumed hydrogen density of $n_{\rm H} = 30 $ cm$^{-3}$ and an escape fraction of $0.4$. This is based on studies of resolved galaxies such as \citet{choi_2020} and \citet{della_bruna_2021} which have estimated the escape fraction for H\scalebox{0.9}{ II} regions to be anywhere between 25\% to 65\%. We add dust in our \textsc{cloudy} model as graphite and silicate grains with size distribution and abundance appropriate for those along the line of sight to the Trapezium stars in Orion, which are then scaled based on the total metallicity of the star particle. 

While the density and escape fraction remain fixed for H\scalebox{0.9}{ II} regions in our models, the size and ionization parameter for each H\scalebox{0.9}{ II} region are allowed vary between each star particle in the following way. The inner radius is set to be the Str\"{o}mgren sphere radius (R$_S$) given by the following equation

\begin{equation}\label{rs_eqn}
R_S = {\Big(\frac{3Q}{4\pi\,n{_H}^2\,\alpha_B}\Big)}^{1/3}    
\end{equation}
Where, Q is the rate of ionizing photons, n${_H}$ is the hydrogen density, and $\alpha_B$ is the case B recombination coefficient at $10^4$ K which is $\alpha_B = 2.5 \times 10^{-13}$ cm$^3$ s$^{-1}$. We assume an escape fraction of 0.4 and the \textsc{cloudy} calculation stops when either the temperature drops below 100 K or if the ionizing fraction drops below 0.1.

\paragraph{Post-AGB stars}
Post-asymptotic giant branch or Post-AGB stars are stars with initial masses between 0.8 and 8 M$_{\odot}$ that have left the asymptotic giant branch and are evolving horizontally along the Hertzsprung-Russell Diagram. While relatively short-lived, post-AGB stars can become hot enough to ionize their ejected outer envelope creating planetary nebulae. \citet{byler_2019} showed that post-AGB stars can be an important ionizing source for the large amounts of diffuse ionized gas emission seen in (.e.g) early-type galaxies. As discussed in section \ref{sec:relative_cont} we find that though post-AGB by themselves do not contribute significantly to the total line emission budget they do act as an important ionizing source for the diffused ionized gas in the galaxy. For modeling post-AGB stars we follow the same prescription as what was used for H\scalebox{0.9}{ II} regions with few minor changes. The gas ionized by the post-AGB stars is primarily the ejected outer envelopes of the parent star which is composed of gas that was enriched by the stellar nucleosynthesis over the lifetime of the star. Thus, we can no longer assume that the stellar metallicity is equal to the gas metallicity in the case of post-AGB stars. 

\citet{stasinska_1998}, \citet{stanghellini_2005}, \citet{stanghellini_2009} and, \citet{henry_2018} have analyzed planetary nebulae in Milky Way Galaxy, Large Magellanic Cloud (LMC) and, Small Magellanic Cloud (SMC) and found that their log(N/O) and log(C/O) was elevated when compared with local observations of H\scalebox{0.9}{ II} regions by about 0.8 and 0.6 dex on average respectively. Taking a conservative approach, to account for the enrichment we enhance both the log(C/O) and log(N/O) for post-AGB stars by 0.4 dex in our model. We also assume a higher escape fraction of 0.6 for post-AGB stars to take into account the fracturing of the gas surrounding the star owing to the stellar evolution.  

\paragraph{Diffuse Ionized Gas (DIG)}\label{sec:dig_method}
The diffuse ionized gas (DIG) component of the ISM can be a major contributor to emission lines fluxes \citep{sanders2017biases,vale2019}. The ionizing radiation from sources such as H\scalebox{0.9}{ II} regions and post-AGB stars can leak out and ionize the surrounding diffuse ISM. To accurately model DIG we need an estimate of the escape fraction of ionizing photons. Recent studies of resolved galaxies such as \citet{choi_2020} and \citet{della_bruna_2021} have estimated the escape fraction for H\scalebox{0.9}{ II} regions to be anywhere between 25\% to 65\%. We assume a constant escape fraction of 40\% for H\scalebox{0.9}{ II} regions and an escape fraction of 60\% post-AGB stars to take into account the potential fracturing of ISM over time due to stellar evolution. 

While we would ideally compute the ionization state in DIG by employing bona fide ionizing radiative transfer, this is computationally intractable for the large numbers of galaxies that we aim to simulate in a cosmological context. As a result, we approximate the emission from DIG in the following manner: we first model the shape of the incident ionizing radiation field from nearby ionizing sources (H\scalebox{0.9}{ II} regions and post-AGB stars) and then scale the flux from this SED to the total energy deposited in the cell from dust radiative transfer. In detail, we first compute the emitting SED from nearby H\scalebox{0.9}{ II} regions and post-AGB stars (within $1$ kpc) using the aforementioned {\sc cloudy} calculations. With this, we compute the ionization parameter via:
\begin{equation}\label{u_eqn}
U = \frac{Q/6}{4\pi\,n{_H}\,c{_W}^2\,c}
\end{equation}
Here, Q is the rate of ionizing photons in $s^{-1}$ incident on the cell. Since there are six faces in a cubic cell, we divide this number by 6 to get the rate of ionizing photons striking each gas cell surface. c${_W}$ is the cell width, c is the speed of light, and n${_H}$ is the hydrogen density which is kept fixed at 10 cm${^3}$ for all the gas cells across redshifts. To compute Q we run $3D$ radiative transfer with the dust radiative transfer code \textsc{hyperion} \footnote{\url{http://www.hyperion-rt.org/}} \citep{Hyperion} (described in more detail shortly). {\sc hyperion} only computes the radiative transfer longward of the Lyman Limit, so we use this calculation simply to compute the total energy deposited in every cell (which is dominated by FUV photons close to the Lyman Limit), and scale the shape of the incident SED from the {\sc cloudy} calculations to match this energy. We only run \textsc{cloudy} on those cells where logU is greater than -6.0 to minimize the computation time. The metallicity of the \textsc{cloudy} model is assumed to be the same as that of the gas cell.  While this is an approximation to the true incident ionizing flux in a cell of diffuse ISM, fully computing the ionizing radiative transfer in large cosmological volumes for numerous snapshots is computationally intractable.

\begin{figure*}[htp]
\centering
\includegraphics[width=\textwidth]{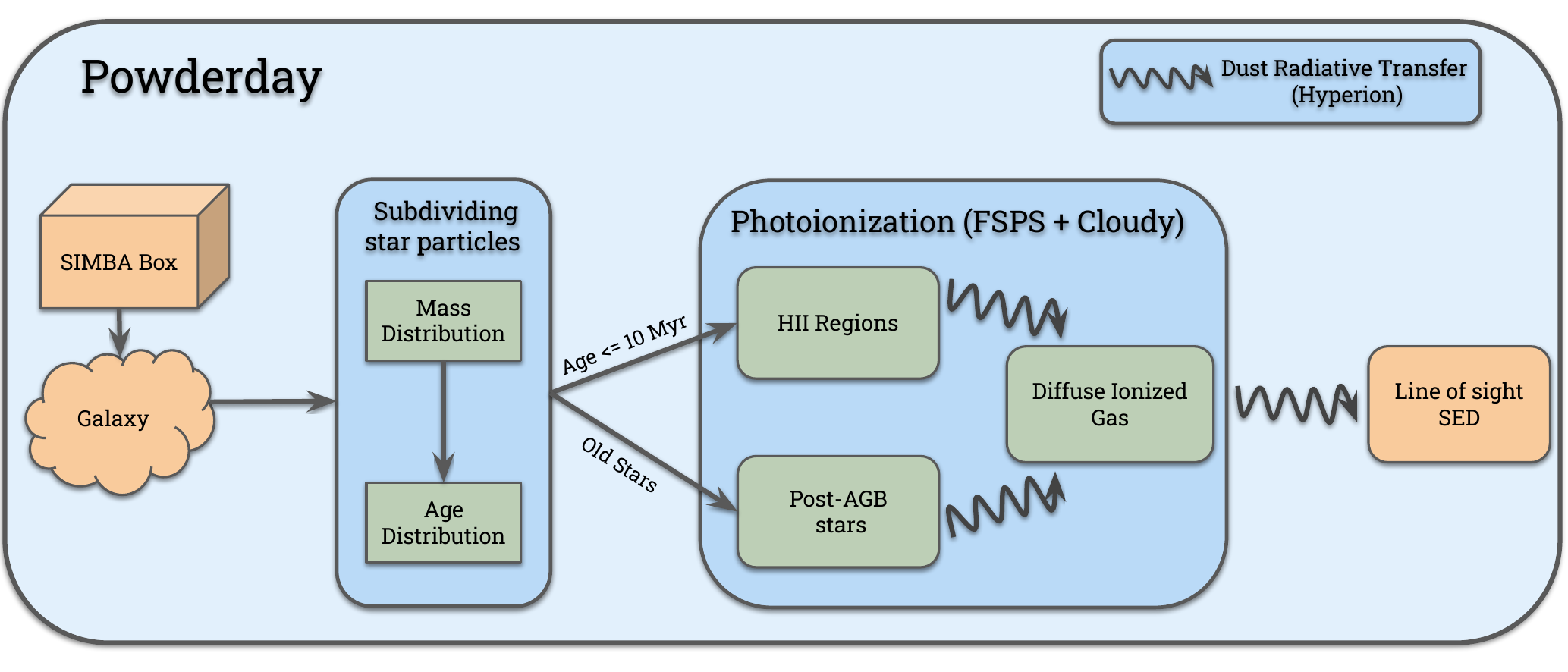}
  \caption{Flowchart showing how our nebular emission modeling pipeline integrated within \textsc{powderday} works. The first step is to identify galaxies within the \textsc{simba} box using \textsc{caesar} galaxy catalog generator. We then subdivide the star particles into a sub-resolution mass and age distribution. Star particles younger than 10 Myr are modeled as an H\scalebox{0.9}{ II} region and old stars are modeled as post-AGB stars. We then run the first pass of dust radiative transfer using \textsc{hyperion} to get the energy deposited in each cell and calculate the ionizing radiation flux for each gas particle within 1 kpc of an H\scalebox{0.9}{ II} region or post-AGB star. We then apply a filter and only model gas cells with sufficiently high ionizing radiation as diffuse ionized gas. After this, a second pass of the dust radiative transfer is performed giving us the line of sight SED with nebular emission for each galaxy.}
  \label{fig:model_diagram}
\end{figure*}

In detail, we have packaged all of the above processes into the radiative transfer package {\sc powderday} \citep{powderday} \footnote{\url{https://powderday.readthedocs.io/en/latest/}}, which acts as a high-level wrapper to calculate nebular emission. The general approach that we use is shown via a flowchart in Figure \ref{fig:model_diagram}. We first run the \textsc{simba} snapshots through \textsc{caesar} galaxy catalog generator. We filter out those galaxies that do not have at least one-star particle younger than 10 Myr. This is done to ensure that we are only considering star-forming galaxies for this study. Then for each galaxy, we take all the star particles less than 10 Myr old and subdivide them based on a cluster mass and age distribution are then assigned a simple stellar population using \textsc{python-fsps}\footnote{\url{http://dfm.io/python-fsps/current/}} \citep{python_fsps} which is a python wrapper for \textsc{fsps}. Each of the subdivided star particles is modeled as a H\scalebox{0.9}{ II} region with a spherical shell geometry and a constant density. Once we are done modeling nebular emission from H\scalebox{0.9}{ II} region we move on to post-AGB stars. For this, we take all the star particles older than 1 Gyr for z = 0, 1, and 100 Myr for z = 2 and above and break them down using a cluster mass and age distribution function. These age limits are used to minimize computation time all the while ensuring we capture as much of the old stellar population as possible. These subdivided star particles are then modeled as post-AGB stars as described above. Once we have finished running the \textsc{cloudy} models for all the H\scalebox{0.9}{ II} regions and post-AGB stars we then use \textsc{hyperion} to do the dust radiative transfer calculation and this gives us the total energy deposited in each gas cell. We then calculate the shape of the ionizing radiation field for each gas cell by taking a distance weighted average of the output \textsc{cloudy} spectrum of all the H\scalebox{0.9}{ II} regions and post-AGB stars that lie within 1 kpc. We filter out gas cells where the log of the ionization parameter is less than -6.0 to minimize computation and then run \textsc{cloudy} on the remaining gas cells. After adding in the DIG sources we perform the final dust radiative transfer calculation and that gives us the line of sight SED with nebular emission for each galaxy in our sample. Finally, the line fluxes are extracted by fitting Gaussian profiles. We calculate the metallicity of each galaxy as the [O\scalebox{0.9}{ III}]$\lambda 5007$ weighted average of the stellar metallicity of all the star particles that emit nebular emission. Throughout this paper, unless otherwise mentioned we report metallicity in units of 12 +log(O/H). The notation that we adopt for the various emission line ratios used in this paper is as follows:

\begin{itemize}
    \item[\textbf{N2}:] log([N\scalebox{0.9}{ II}]$\lambda 6583$/H$\alpha$)
    \item[\textbf{R3}:] log([O\scalebox{0.9}{ III}]$\lambda 5007$/H$\beta$)
    \item[\textbf{R2}:] log([O\scalebox{0.9}{ II}]$\lambda 3727, 29$/H$\beta$)
    \item[\textbf{R23}:] log(([O\scalebox{0.9}{ II}]$\lambda 3727, 29$ + [O\scalebox{0.9}{ III}]$\lambda 4959, 5007)$/H$\beta$)
    \item[\textbf{O3N2}:] log(([O\scalebox{0.9}{ III}]$\lambda 5007$/H$\beta$) /  ([N\scalebox{0.9}{ II}]$\lambda 6583$/H$\alpha$))
    \item[\textbf{R3N2}:] log([O\scalebox{0.9}{ III}]$\lambda 5007$ / [N\scalebox{0.9}{ II}]$\lambda 6583$)
    \item[\textbf{N2O2}:] log([N\scalebox{0.9}{ II}]$\lambda 6583$ / [O\scalebox{0.9}{ II}]$\lambda 3727$)
    \item[\textbf{O3O2}:] log([O\scalebox{0.9}{ III}]$\lambda 5007$ / [O\scalebox{0.9}{ II}]$\lambda 3727, 29$)
    \item[\textbf{Ne3O2}:] log([Ne\scalebox{0.9}{ III}]$\lambda 3869$ / [O\scalebox{0.9}{ II}]$\lambda 3727, 29$)
    \item[\textbf{S2}:] log([S\scalebox{0.9}{ II}]$\lambda 6717, 31$/H$\alpha$)
    \item[\textbf{O3S2}:] log(([O\scalebox{0.9}{ III}]$\lambda 5007$/H$\beta$) / ([S\scalebox{0.9}{ II}]$\lambda 6717, 31$/H$\alpha$))
    \item[\textbf{N2S2}:] log(([N\scalebox{0.9}{ II}]$\lambda 6583$/[S\scalebox{0.9}{ II}]$\lambda 6717, 31$) \\ + 0.264([N\scalebox{0.9}{ II}]$\lambda 6583$/H$\alpha$))  
\end{itemize}

\vspace{5mm}
\section{Local Strong line metallicity calibrations}\label{sec:local_mi}
\begin{figure*}
\centering
  \includegraphics[width=\textwidth]{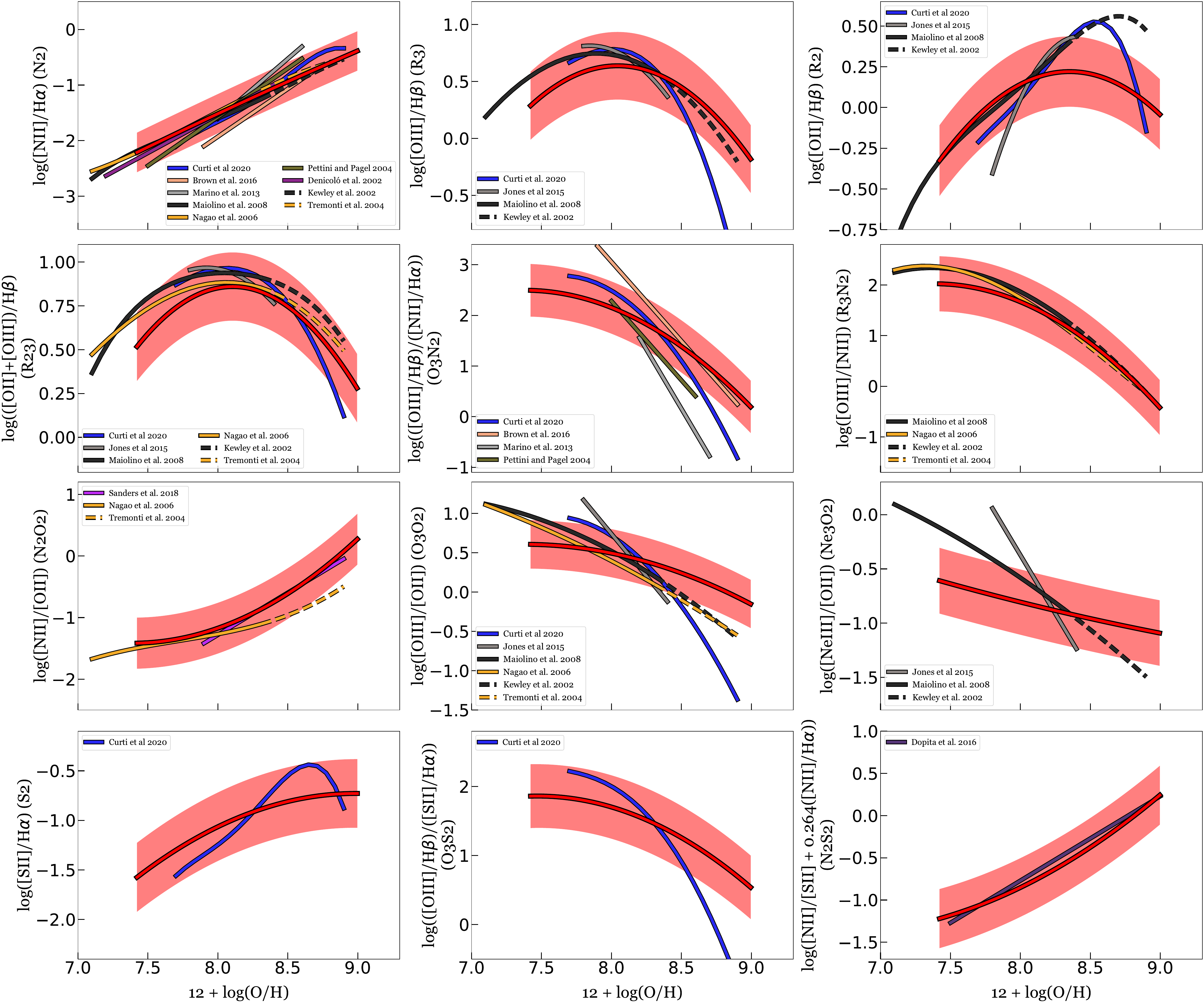}
  \caption{A comparison between the theoretical strong line metallicity calibrations for the 12 line ratios using our model on the \textsc{simba} z = 0 galaxies against different local observational and theoretical calibrations from the literature \citep{denicolo_2002, kewley2002, tremonti2004, nagao_2006, Pettini, nagao_2006, Maiolino2008, marino_2013, Jones2015, brown_2016, dopita2016, sanders_mosdef_2018, curti_2020}. In all the panels the thick red line shows the calibration from \textsc{simba} and the red shaded region shows the 90\% confidence interval. We fit polynomials of the form $y = a_2*x^2 + a_1*x^1 + a_0$ where x is the metallicity (12 + log(O/H)) and y is the line ratio. The metallicity calibrations from the literature that were calibrated using auroral line measurements are shown by solid colored lines whereas the dashed lines represent the theoretical strong line calibrations that were used to extend them in the regime where there was no observational data available.}
  \label{fig:z0_all}
\end{figure*}

Before we can use our model to extend the strong line metallicity calibrations to high redshift galaxies we first ensure that our model can accurately reproduce the locally observed auroral line-based metallicity calibrations from the literature. In Figure \ref{fig:z0_all}, we compare the theoretical strong line metallicity calibrations derived using \textsc{simba} $z=0$ snapshot against different local observational and theoretical calibrations from the literature. We consider all $12$ modeled line ratios and perform a fit third-degree polynomial of the form $y = a_2*x^2 + a_1*x^1 + a_0$ where x is the metallicity (12 + log(O/H)) and y is the line ratio, to the data using ordinary least square regression represented by the thick red line. The red shaded area represents the region of $90 \%$ confidence interval, while the other solid color lines are the observational auroral line-based strong line calibrations. The dashed lines are theoretical calibrations that were used to extend the auroral line calibrations in the regime where no observational data was present. As can be seen, our model provides a reasonable match to the observed calibrations for the majority of the line ratios. We now discuss the individual line ratios in turn:

\paragraph{N2}
We compare our calibration fit for N2 against the calibrations from \citet{denicolo_2002, kewley2002, tremonti2004, Pettini, nagao_2006, Maiolino2008, marino_2013, brown_2016, curti_2020}. As can be seen in Figure \ref{fig:z0_all}, our model performs well in matching the calibrations from the literature across the entire metallicity range. The nature of the curve can be understood by the fact that as the metallicity decreases, the H\scalebox{0.9}{ II} region temperature rises due to a decrease in metal cooling. This causes more of the nitrogen to occupy higher ionization states. This, coupled with the decreasing abundance of nitrogen with decreasing metallicity leads to the [N\scalebox{0.9}{ II}]$\lambda 6583$ being a monotonically decreasing function of metallicity. 

\paragraph{R3, R2 and R23}
These line ratios are all double-valued functions of metallicity. This is because, at high metallicities, temperatures become too to collisionally excite the valence electrons of oxygen ions into higher electron energy levels. As the metallicity decreases, the increase in ionization parameter and changes in the ionization spectral shape/hardness leads to more O++ emission, and as a result, the line ratios increase. We can see that [O\scalebox{0.9}{ II}]$\lambda 3727, 29 $/H$\beta$ reaches a peak around metallicity of 12 + log(O/H) $\sim$ 8.5 whereas [O\scalebox{0.9}{ III}]$\lambda 5007 $/H$\beta$ reaches a peak at a lower metallicity of 12 + log(O/H) $\sim$ 8.0 due to it being a doubly ionized species. As we move to lower metallicities, the decrease in the amount of oxygen available to excite becomes the primary limiting factor and we see that the line ratios start to decrease, giving rise to the double-valued shape. In Figure \ref{fig:z0_all} we compare our calibration for R2, R3 and R23 against calibrations from \citet{kewley2002, tremonti2004, nagao_2006, Maiolino2008, Jones2015} and \citet{curti_2020}. In general, for R3 and R23 our calibrations underestimate the line ratio at lower metallicities (12 + log(O/H) $<$ 8) but the overall nature of the curve is similar to the literature calibrations. As for R2, our calibrations work well at low metallicities but are unable to reach high enough values at high metallicities (12 + log(O/H) $>$ 8). As discussed in \ref{sec:relative_cont} (Figure \ref{fig:rc}) we find that diffuse ionized gas is a substantial contributor to the total line emission budget of [O\scalebox{0.9}{ II}]$\lambda 3727, 29 $ at high metallicities (12 + log(O/H) $>$ 8) accounting for as much as 70\% of the total line emission budget. As such the mismatch at high metallicities might be explained by the uncertainties in our DIG modeling. 

\paragraph{O3N2 and R3N2}
We compare our calibration for O3N2 and R3N2 against the observed auroral line and theoretical calibrations from \citet{Pettini, kewley2002, tremonti2004, nagao_2006, Maiolino2008, marino_2013, brown_2016, curti_2020}. We find that for R3N2 our calibrations are an excellent match to the observations whereas for O3N2 we do a reasonable job at low metallicities but at high metallicities (12 + log(O/H) $>$ 8.5), we tend to somewhat overestimate the line ratio. That said, there is significant uncertainty in the observational calibrations themselves. Both of these calibrations are decreasing functions of metallicity with decreasing slopes at lower metallicities. This is because as we move to metallicities lower than $\sim 8.0$ in units of 12+log(O/H), the [O\scalebox{0.9}{ III}]$\lambda 5007 $ starts to decrease which causes the curve to flatten out and to even eventually move to a downward trend at even lower metallicities.

\paragraph{N2O2}
We compare our calibration for N2O2 against calibrations from \citet{tremonti2004, nagao_2006, sanders_mosdef_2018}. Note, that the two observed calibrations do not agree among themselves because \citet{sanders_mosdef_2018} corrected for DIG contribution whereas \citet{nagao_2006} did not. As a result of this, the calibration from \citet{nagao_2006} flattens out at low metallicity whereas there is no such flattening observed in the calibration from \citet{sanders_mosdef_2018}. On the other hand, at high metallicities the calibration from \citet{sanders_mosdef_2018} has a much higher slope than the calibration from \citet{nagao_2006}. We find that on the high metallicity end, we find a close match to the calibrations from \citet{sanders_mosdef_2018}. We would like to note that it is somewhat surprising that our models that include DIG provide such a good match to \citet{sanders_mosdef_2018} which as mentioned above corrected the SDSS calibration by removing the contribution from DIG. On the low metallicity end (12 + log(O/H) $<$ 8), we find that our model closely matches the calibrations presented in \citet{nagao_2006}. This resemblance is observed in the flattening of the curve, as the [OII] emission line exhibits a turning point and begins to decrease at lower metallicities.

\paragraph{O3O2 and Ne3O2}
We compare our calibrations for O3O2 and Ne3O2 against calibrations from the \citet{tremonti2004, kewley2002, nagao_2006, Maiolino2008, Jones2015, curti_2020}. As stated earlier, we tend to underestimate R3 at low metallicities (12 + log(O/H) $<$ 8), and at high metallicities (12 + log(O/H) $>$ 8.5) underestimate R2. The net result of this is that we find that our calibration for O3O2 is flatter. Similarly, we find that our calibration for Ne3O2 also has a lower slope than the observed calibrations. That said, there is a large variation in slope across different observed calibrations for both the line ratios as well though none are as flat as the simulations. This flatness can be due to the fact that though we find that an anti-correlation relation between logU and O/H exists in our models it is a lot flatter than what is reported by other studies like \citet{perez-montero_2014}.

\paragraph{S2 and O3S2}
We find that our calibration for S2 matches the observed calibration from \citet{curti_2020} fairly well at low metallicities (12+ log(O/H) < 8.2). The overall nature of the curve is much flatter than the observed calibration which means at high metallicities (12+ log(O/H) > 8.2) our calibration does not reach values as high as what is observed. Similarly, for O3S2 we find that our calibration has the same general nature with the line ratio increases with decreasing metallicity albeit with a flatter slope which can be attributed to the underestimation of [S\scalebox{0.9}{ II}]$\lambda 6717, 31$ at high metallicities. The mismatch at high metallicities might be explained by the uncertainties in modeling DIG emission (see section \ref{sec:caveats} for more info). Similar to [O\scalebox{0.9}{ II}]$\lambda 3727, 29 $ we find that diffuse ionized gas can account for as much as 70 \% of the total ionizing flux of [S\scalebox{0.9}{ II}]$\lambda 6717, 31$ at high metallicities (Figure \ref{fig:rc}).
 
\paragraph{N2S2}
In the last panel of Figure \ref{fig:z0_all} we compare our calibration for N2S2 against calibration from \citet{dopita2016}. We can see that our model performs reasonably well at matching the observed calibration, although we do see slight deviations at the extreme ends. 
\newline
\newline
In summary, we find that our model does a fairly reasonable job reproducing the general nature of the curve for most of the line ratios. We can match the observed auroral line calibrations fairly well at low metallicities. However, for line ratios like R2, S2, and O3S2 that involve singly ionized oxygen and sulfur lines, our model under-performs at high metallicities (12 + log(O/H) $>$ 8) and the fact that diffuse ionized gas dominates the line emission budget for these lines at high metallicities coupled with the fact that our current model is unable to accurately capture that emission might explain the mismatch.

\section{Strong line Metallicity calibrations across redshifts}\label{sec:z_mi}
\begin{figure*}
\centering
  \includegraphics[width=\textwidth]{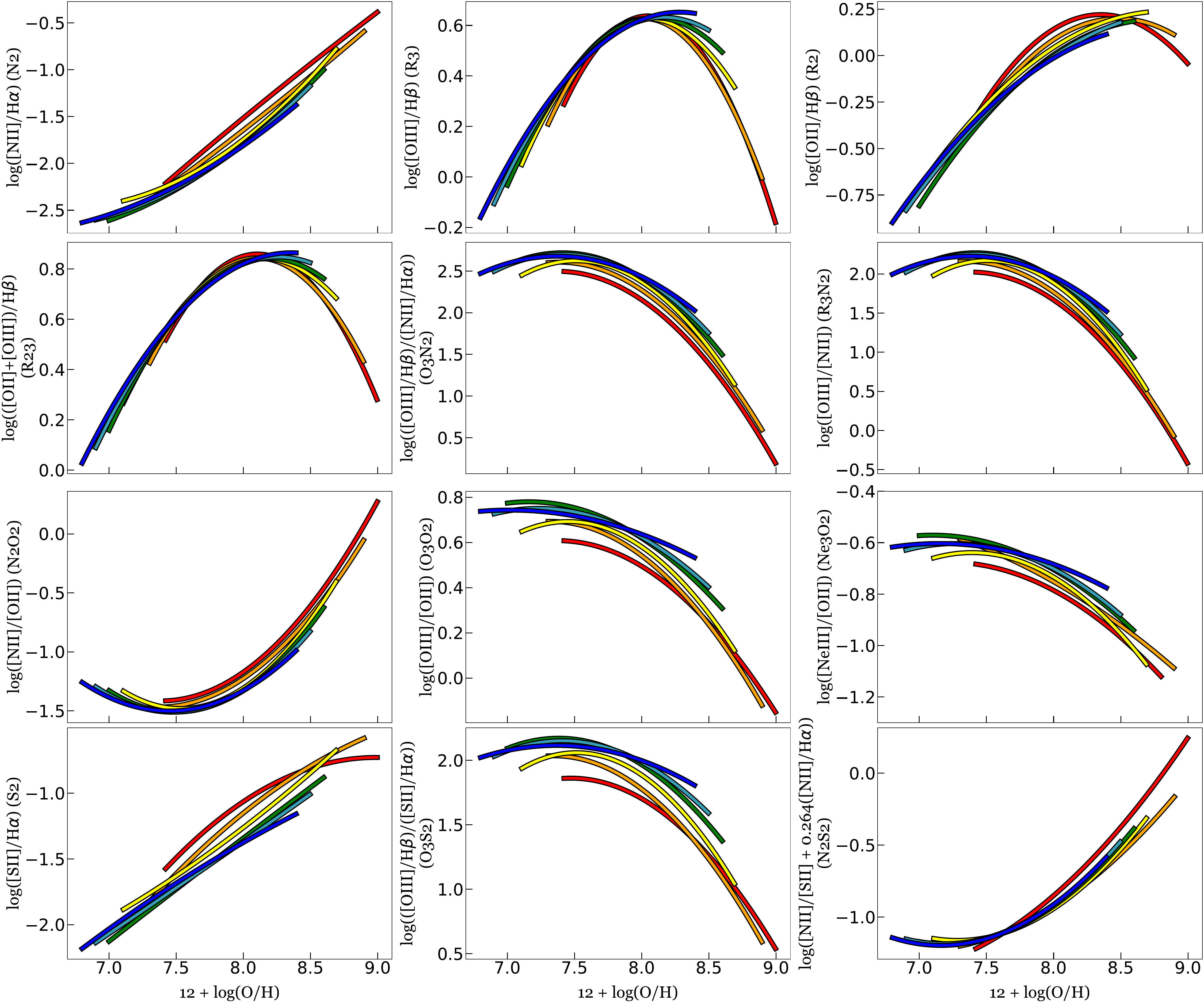}
  \caption{The redshift evolution of the strong line metallicity calibrations using our nebular emission model. For the 12 line ratios, the calibrations going from z = 0 to z = 5 are shown by different colored lines going from red to blue with increasing redshift. We fit polynomials of the form $y = a_2*x^2 + a_1*x^1 + a_0$ where x is the metallicity (12 + log(O/H)) and y is the line ratio. We find that N2, O3N2, R3N2, N2O2, O3O2, Ne3O2, S2 and O3S2 line ratios show mild evolution of about 0.2 - 0.4 dex with redshift. In contrast, indicators like R2, R3, R32 and N2S2 show little redshift evolution. }
  \label{fig:z_all}
\end{figure*}
In the previous section, we discussed how our metallicity calibrations fare against redshift $z\sim0$ constraints anchored in auroral line-based metallicity indicators. We now proceed in applying our model to high redshift galaxies in order to study and understand the redshift evolution of strong line metallicity indicators. We remind the reader that the elements of our model that vary with redshift are: 
\begin{itemize}
    \item Stellar and gas metallicities.
    \item Spectrum of ionizing photons in H\scalebox{0.9}{ II} regions and post-AGB stars.
    \item Incident radiation field for calculating DIG emission.
    \item Number of young and post-AGB stars.
    \item Amount of gas available for ionization.
\end{itemize}

In Figure \ref{fig:z_all} we plot the 12 line ratios we consider at integer redshifts from $z=0-5$. In short, we find mild redshift evolution for all of the line ratios that we model, aside from the fact that the majority of high-$z$  galaxies lie in a lower metallicity regime of each line ratio metallicity diagnostic. This suggests that the change in stellar and gas metallicities, incident radiation fields, and the number of star and gas particles with redshift do have a discernible effect on the strong line calibrations causing them to vary from low to high redshift in our model. Recent JWST observations of auroral lines galaxies spanning redshift range from $\sim$ 2 - 9 \citep{curti_2022, trump_2022, nakajima_2023, sanders_direct_2023} also indicate that the metallicity indicators might evolve with redshift. We discuss this in detail along with the uncertainties and caveats of our model in section \ref{sec:discussion}. 

An important point to note is that the utility of the line ratios changes depending on what metallicity range or redshift one is interested in. This in turn affects the applicability of the line ratios across redshifts. For example, the line ratio N2S2 (Figure \ref{fig:z_all}; last panel) is a monotonically increasing function of metallicity at high metallicity / low redshifts but becomes double valued function of metallicity at low metallicities / high redshifts. Thus, a line ratio that works well at low redshifts might not be an appropriate choice for high redshift galaxies. In the following subsections, we will look at the 12-line ratios in detail and explore the general nature of the curve. We will mainly focus on exploring the low and high metallicity regimes and discussing where different metallicity indicators work well and the regimes where they must be avoided.

\subsection{Discussion of Current Standard Calibrators}
\paragraph{N2}
We find that logN2 varies by about 0.2 - 0.4 dex from redshift 0 to 5 and the variation decreases with metallicity. However, we would like to point out that this variation is smaller than the 90\% confidence intervals at z = 0 that span a width of about 0.7 dex (see Figure \ref{fig:z0_all}). N2 is a reasonably robust metallicity indicator across all redshifts. N2 is a single-valued function with a steep slope which makes it an excellent indicator over the whole metallicity range except at low metallicities (12 + log(O/H) < 7.0) where the slope starts to flatten out. Nonetheless, at low metallicities (common at high-z), it is very challenging to detect [N\scalebox{0.9}{ II}]$\lambda 6583$ and it can be even weaker than [O\scalebox{0.9}{ III}]$\lambda 4363$ auroral line in many cases at these redshifts. Thus, in practice not many very high-z galaxies (z>4) have [N\scalebox{0.9}{ II}]$\lambda 6583$ detections in JWST spectroscopy, which limits its utility (see \citet{sanders_2023} and \citet{shapley_2023b} for more info). Further, it is important to recall a significant caveat concerning line ratios involving [N\scalebox{0.9}{ II}]$\lambda 6583$ in our model. \textsc{simba} is unable to reproduce the observed log(N/O) vs log(O/H) relation in nearby galaxies \citep{pilyugin2012,garg2022bpt}. Therefore, in our model, we fix the nitrogen abundances according to Equation \ref{pilyugin_eqn} to match the observed abundances. If it turns out that the log(N/O) vs log(O/H) relation evolves with redshift then that will introduce uncertainty in our calibrations involving [N\scalebox{0.9}{ II}]$\lambda 6583$ at high redshifts (see \secref{sec:method_sources} and \secref{sec:caveats} for more info).

\paragraph{R3, R2 and R23}
R3, R2, and R23 all involve oxygen emission lines and behave similarly, we find almost no evolution in any of these metallicity indicators across redshifts. [O\scalebox{0.9}{ III}]$\lambda 5007$ is one of the brightest lines in the spectrum making it easy to observe even in faraway galaxies which makes these ratios one of the most widely accessible across redshifts. Unfortunately, we find these line ratios, especially R3 and R23, to be quite poor metallicity calibrations and should be used only if a better alternative is not available, especially at low redshifts. This is because at metallicities above 7.0 in units of 12+log(O/H), these are double values functions and at the same value of line ratio one can have two solutions for metallicity. Therefore, in the absence of any other constraint on metallicity, like stellar mass, there is no way to know which value is correct. However, as we go to lower metallicities or higher redshifts, these line ratios eventually do become single-valued and can be used as metallicity indicators reasonably well.

\paragraph{O3N2, R3N2 and N2O2}
We find that O3N2, R3N2, and N2O2 all show mild variations of about 0.2 -0.3 dex with redshift. O3N2 and R3N2 increase at the same metallicity with redshift whereas N2O2 decreases with redshift at the same metallicity. We find that they are all reasonable metallicity indicators at high metallicities / low redshifts with an average 90\% confidence interval spread of about 0.7 dex at redshift 0 which is greater than the variation seen across redshift. As we move to the low metallicity regime their utility becomes fairly limited. Below 12 + log(O/H) $\approx$ 8.0 the curves start to flatten out and eventually start to turn over and become double valued which limits the scenarios in which these metallicity indicators can be utilized.

\paragraph{O3O2 and Ne3O2}
In general, we find that O3O2 and Ne3O2 show an increase of about 0.2 dex with increasing redshift at the same metallicities. Both the indicators are a monotonic function of metallicity in the low redshifts/high metallicity regime while they become double valued at metallicities below 7.5 in units of 12 + log(O/H) as the curve starts to flatten out. Both these line ratios have a fairly wide 90\% confidence interval which coupled with a low slope hampers the usefulness as a metallicity indicator at all but extremely high metallicities.

\paragraph{S2}
We find that S2 decreases by about 0.4 dex between z = 0 and z = 5 at the same metallicities. In general, we find S2 to be an excellent metallicity indicator across the whole metallicity range except at very high metallicities (12 + log(O/H) > 8.5) where the curve starts to flatten out. That said, due to the significant contribution of the diffuse ionized gas to the total line flux at high metallicities coupled with the uncertainties involved with our modeling of DIG one should be careful when using S2 at very high metallicities (see \secref{sec:relative_cont} for more info). S2 is a monotonically increasing function of metallicity and has a fairly high and almost constant slope with 0.8 dex a 90 \% confidence interval.

\paragraph{O3S2 and N2S2}
We find that O3S2 increases by about 0.3 - 0.4 dex from z = 0 to z = 5 at the same metallicity whereas we find almost no evolution in N2S2 except at very high metallicities. Both O3S2 and N2S2 metallicity indicators work well at high metallicities (12 + log(O/H) > 8.0) but metallicities lower than 8.0, they are double-valued functions of metallicity and unless there is a way to break the degeneracy this limits their application as metallicity indicators in the low metallicity/high redshift regime.

\onecolumngrid  
\begin{longtblr}[
  caption = {Theoretical strong line calibrations using our model at z = 0 for different line ratios. We fit polynomials of the form $y = a_2*x^2 + a_1*x^1 + a_0$ where x is the metallicity (12 + log(O/H)) and y is the line ratio. For fits to higher redshift galaxies see Appendix \ref{app:mi_higZ}.},
  label = {tab:mi_z0},
]
{
  colspec = {|c|c|c|c|c|c|},
  rowhead = 1,
  row{2-5} = {m, cyan9},
  row{6-9} = {m},
  row{10-13} = {m, cyan9},
  row{14-17} = {m},
  row{18-21} = {m, cyan9},
  row{22-25} = {m},
  row{26-29} = {m, cyan9},
  row{30-33} = {m},
  row{34-37} = {m, cyan9},
  row{38-41} = {m},
  row{42-45} = {m, cyan9},
  row{46-49} = {m},
  row{1} = {olive9}
  }
\hline
\textbf{Line Ratio} & \textbf{Fit} & \textbf{Lower CI} & \textbf{Upper CI} & \textbf{Metallicity Range} \\
\hline
\textbf{N2} & {$a_0$: -13.9 \\ $a_1$: 1.918 \\ $a_2$: -0.046} & {$a_0$: -13.417 \\ $a_1$: 1.887 \\ $a_2$: -0.044} & {$a_0$: -14.382 \\ $a_1$: 1.949 \\ $a_2$: -0.048} & 7.4 - 9.0\\ 
\hline
\textbf{R3} & {$a_0$: -57.889 \\ $a_1$: 14.552 \\ $a_2$: -0.905} & {$a_0$: -57.482 \\ $a_1$: 14.525 \\ $a_2$: -0.903} & {$a_0$: -58.296 \\ $a_1$: 14.578 \\ $a_2$: -0.906} & 7.4 - 9.0\\ 
\hline
\textbf{R2} & {$a_0$: -43.87 \\ $a_1$: 10.558 \\ $a_2$: -0.632} & {$a_0$: -43.574 \\ $a_1$: 10.539 \\ $a_2$: -0.631} & {$a_0$: -44.166 \\ $a_1$: 10.577 \\ $a_2$: -0.633} & 7.4 - 9.0\\ 
\hline
\textbf{R23} & {$a_0$: -47.251 \\ $a_1$: 11.872 \\ $a_2$: -0.732} & {$a_0$: -46.982 \\ $a_1$: 11.854 \\ $a_2$: -0.731} & {$a_0$: -47.521 \\ $a_1$: 11.89 \\ $a_2$: -0.733} & 7.4 - 9.0\\ 
\hline
\textbf{O3N2} & {$a_0$: -45.251 \\ $a_1$: 12.947 \\ $a_2$: -0.878} & {$a_0$: -44.543 \\ $a_1$: 12.902 \\ $a_2$: -0.875} & {$a_0$: -45.959 \\ $a_1$: 12.993 \\ $a_2$: -0.88} & 7.4 - 9.0\\ 
\hline
\textbf{R3N2} & {$a_0$: -49.279 \\ $a_1$: 13.895 \\ $a_2$: -0.941} & {$a_0$: -48.537 \\ $a_1$: 13.847 \\ $a_2$: -0.938} & {$a_0$: -50.021 \\ $a_1$: 13.944 \\ $a_2$: -0.944} & 7.4 - 9.0\\ 
\hline
\textbf{N2O2} & {$a_0$: 34.201 \\ $a_1$: -9.642 \\ $a_2$: 0.652} & {$a_0$: 34.765 \\ $a_1$: -9.678 \\ $a_2$: 0.655} & {$a_0$: 33.638 \\ $a_1$: -9.605 \\ $a_2$: 0.65} & 7.4 - 9.0\\ 
\hline
\textbf{O3O2} & {$a_0$: -15.091 \\ $a_1$: 4.258 \\ $a_2$: -0.289} & {$a_0$: -14.67 \\ $a_1$: 4.231 \\ $a_2$: -0.287} & {$a_0$: -15.511 \\ $a_1$: 4.285 \\ $a_2$: -0.29} & 7.4 - 9.0\\ 
\hline
\textbf{Ne3O2} & {$a_0$: 3.986 \\ $a_1$: -0.877 \\ $a_2$: 0.035} & {$a_0$: 4.412 \\ $a_1$: -0.907 \\ $a_2$: 0.037} & {$a_0$: 3.559 \\ $a_1$: -0.847 \\ $a_2$: 0.033} & 7.4 - 9.0\\ 
\hline
\textbf{S2} & {$a_0$: -28.699 \\ $a_1$: 6.227 \\ $a_2$: -0.347} & {$a_0$: -28.226 \\ $a_1$: 6.196 \\ $a_2$: -0.345} & {$a_0$: -29.171 \\ $a_1$: 6.257 \\ $a_2$: -0.348} & 7.4 - 9.0\\ 
\hline
\textbf{O3S2} & {$a_0$: -29.935 \\ $a_1$: 8.511 \\ $a_2$: -0.57} & {$a_0$: -29.305 \\ $a_1$: 8.47 \\ $a_2$: -0.567} & {$a_0$: -30.565 \\ $a_1$: 8.552 \\ $a_2$: -0.572} & 7.4 - 9.0\\ 
\hline
\textbf{N2S2} & {$a_0$: 11.406 \\ $a_1$: -3.871 \\ $a_2$: 0.292} & {$a_0$: 11.882 \\ $a_1$: -3.901 \\ $a_2$: 0.294} & {$a_0$: 10.931 \\ $a_1$: -3.84 \\ $a_2$: 0.29} & 7.4 - 9.0\\ 
\hline
\end{longtblr}
\twocolumngrid

In summary, we find that indicators like N2, O3N2, R3N2, N2O2, O3O2, Ne3O2, S2, and O3S2 show mild evolution of about 0.2 - 0.4 dex with redshift. Whereas, indicators like R2, R3, R32, and N2S2 show almost no redshift evolution. We find S2 to be an excellent metallicity indicator across a metallicity range of 6.8 - 8.5 in units of 12+log(O/H) and then it starts to flatten out whereas N2 works fairly well at all metallicities except at extremely low end. Line ratios such as  R2, R3, and R23 work fairly well at low metallicities/high redshifts but become a double-valued function at high metallicities which hampers their utility at low redshifts. On the other hand R3N2, O3N2, N2O2, O3S2, and N2S2 all work fairly well at low redshifts but are double valued at low metallicities limiting their applicability at high redshifts. In contrast to observations relatively shallow slopes and fairly wide confidence intervals hamper the usefulness of O3O2 and Ne3O2 across the whole metallicity range. We recommend using multiple metallicity indicators whenever possible to mitigate the uncertainty. In Table \ref{tab:mi_z0} we list the polynomial fits to all the metallicity indicators discussed above along with the fits for the lower and upper 90\% confidence intervals.

\subsection{Relative contribution of different ionizing sources}\label{sec:relative_cont}
\begin{figure}
\centering
  \includegraphics[width=\columnwidth]{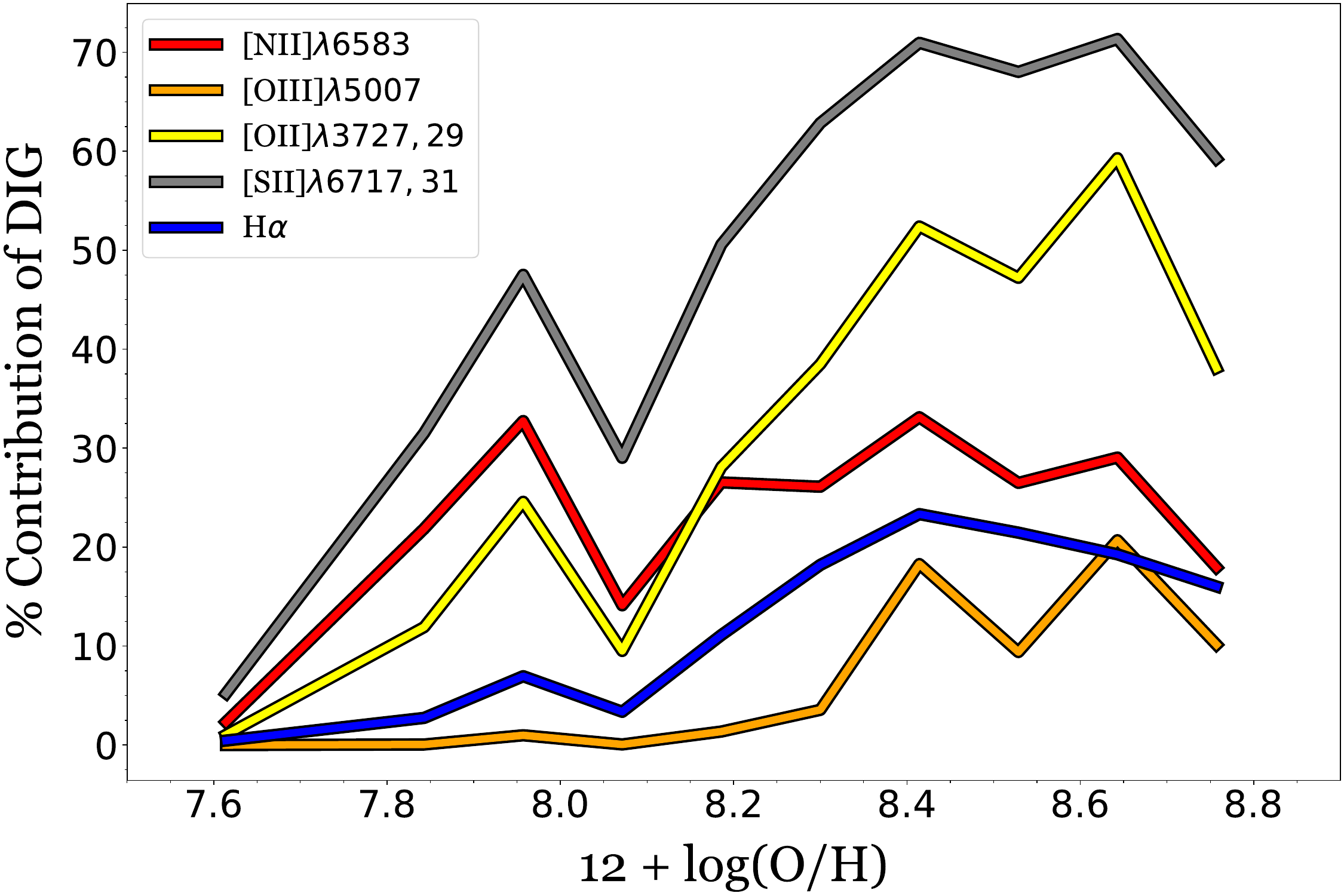}
  \caption{The relative contribution of diffuse ionized gas to the total line flux with respect to metallicity (12 +log(O/H) for \textsc{simba} z = 0 galaxies. DIG contribution for [N\scalebox{0.9}{ II}]$\lambda 6583$, [O\scalebox{0.9}{ III}]$\lambda 5007$, [O\scalebox{0.9}{ II}]$\lambda 3727, 29 $, [S\scalebox{0.9}{ II}]$\lambda 6717, 31$ and H$\alpha$  are shown in red, orange, yellow, gray and blue respectively. We find that DIG is an important source of nebular emission and can account for as much as 70\% of the total line flux in extreme cases.}
  \label{fig:rc}
\end{figure}

In this section, we discuss how the relative contribution of different ionizing sources in a galaxy, namely H\scalebox{0.9}{ II} regions around young stars, Post-AGB stars, and DIG, vary with metallicities for different emission lines. In figure \ref{fig:rc} we plot the percentage of the total line flux that is emitted by diffuse ionized gas for [N\scalebox{0.9}{ II}]$\lambda 6583$, [O\scalebox{0.9}{ III}]$\lambda 5007$, [O\scalebox{0.9}{ II}]$\lambda 3727, 29 $, [S\scalebox{0.9}{ II}]$\lambda 6717, 31$ and H$\alpha$. To simplify interpretation, in Figure \ref{fig:rc} we only plot the percentage contribution of DIG against metallicity at z = 0: In all the cases, we find that the post-AGB stars contribute exceedingly little to the total line emission budget, and therefore ignore their contribution. However, their main contribution comes in the fact that they act as an important source of ionizing radiation for producing diffuse ionized gas. \citet{vale2019} and \citet{sanders2017biases} have argued that diffuse ionized gas can be a substantial contributor and can account for as much as 50\% of the line emission budget in local galaxies. Similarly, we find that DIG gas plays a significant role in generating nebular emission and the contribution of DIG is positively correlated with metallicity for all five lines. For [N\scalebox{0.9}{ II}]$\lambda 6583$, [O\scalebox{0.9}{ III}]$\lambda 5007$ the H\scalebox{0.9}{ II} regions around young stars are the main contributor to the line flux at all metallicities but DIG can still account for as much as 20 - 30\% of the total line flux in some cases. This is true even for H$\alpha$ which is independent of the complex abundance pattern and ionization state issues that affect the forbidden metal lines. In the cases of [O\scalebox{0.9}{ II}]$\lambda 3727, 29 $ and [S\scalebox{0.9}{ II}]$\lambda 6717, 31$ we find that the contribution of DIG can account for as much as 70\% of the total flux at high metallicities. Thus, DIG is an important source of nebular emission and must not be omitted when modeling nebular emission in galaxy-integrated spectra.

We observe a positive correlation between the percentage of DIG contribution to the nebular flux, and metallicity. This may be due to a number of possible reasons. First, this may just be due to galaxies with higher metallicity having larger stellar masses. This would mean that high metallicity galaxies can have more ionizing photons escaping from H\scalebox{0.9}{ II} regions and post-AGB stars that can then ionize the surrounding gas particles, therefore increasing the relative contribution of diffuse ionized gas. On the other hand, there can be a variety of other factors for example, local observations from \citet{oey_2007} suggest that the fractional DIG contribution is related to star formation rate surface density ($\Sigma_SFR$), which may be in turn depend on specific star formation rate (sSFR). Thus, a varying specific sSFR with metallicity, with high-metallicity galaxies tending to have lower sSFR or evolving volume filling fraction in the ISM can also impact the relative contribution of DIG.
\\
\\
\section{Discussion}\label{sec:discussion}
\begin{figure*}
\centering
  \includegraphics[width=\textwidth]{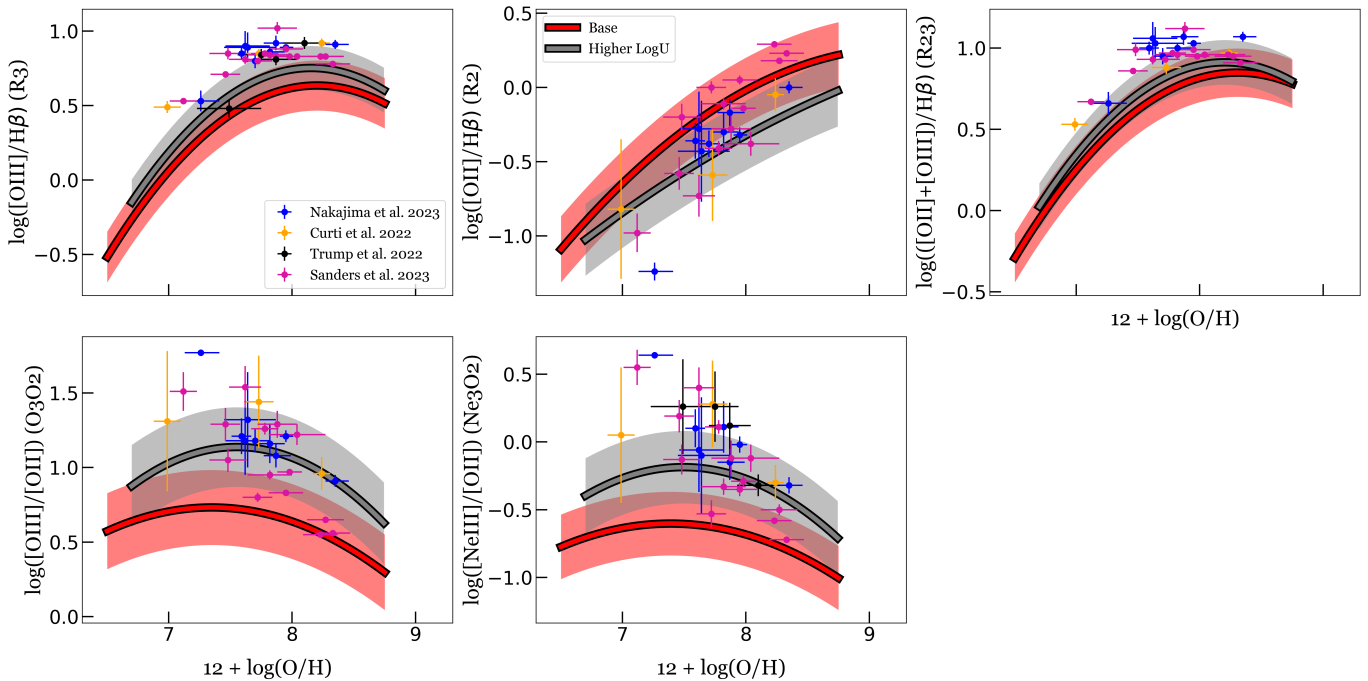}
  \caption{A comparison of our calibrations using \textsc{simba} galaxies with the base model at z = 5 and an increased ionization parameter model at z = 5 against recent JWST observations of high-z galaxies. We fit polynomials of the form $y = a_2*x^2 + a_1*x^1 + a_0$ where x is the metallicity (12 + log(O/H)) and y is the line ratio. We find that the increased ionization parameter model provides a much better match to the observations as compared to our base model implying that the metallicity calibrations might evolve with redshift towards a higher ionization parameter. The observations from \citet{curti_2022, trump_2022, nakajima_2023, sanders_direct_2023} are shown by orange, black, blue, and pink points respectively. In all the panels the red line shows the calibration using our base model and the gray line shows our calibrations for an increased ionization parameter model. The shaded red and gray regions denote the 90\% confidence interval for the two models.}
  \label{fig:jwst_mi_comp}
\end{figure*}
\subsection{Comparison to Other Theoretical Models and Relevant Observations}\label{sec:comparison}
In this section, we review some of the recent observations and theoretical studies that look at the redshift evolution of nebular line ratios. Recently, major JWST programs such as the Cosmic Evolution Early Release Survey (CEERS), GLASS-ERS, and the Advanced Deep Extragalactic Survey (JADES) have observed nebular lines in a number of galaxies spanning a redshift range from z $\sim$ 5 - 9 \citep{curti_2022, cameron_2023, trump_2022, sanders_2023, nakajima_2023, sanders_direct_2023}. Some of these studies \citep{curti_2022, trump_2022, nakajima_2023, sanders_direct_2023, shapley_2023a, shapley_2023b} were sufficiently sensitive to be able to detect the [O\scalebox{0.9}{ III}]$\lambda 4363$ auroral line in approximately 30 galaxies and obtain an estimate of gas phase metallicity using the direct $T_e$ measurement for the first time at such high redshifts. 

The observations span a wide range from z $\sim$ 5 - 9 but our models do not go beyond z = 5. This is not a big issue for R3, R2, and R23 line ratios but we have shown that O3O2 and Ne3O2 do evolve with redshift, showing an increase of about 0.2 dex between z = 0 and z = 5. We plot the observations from \citet{curti_2022, trump_2022, nakajima_2023, sanders_direct_2023} against our model calibrations at z = 5 (red line) in Figure \ref{fig:jwst_mi_comp}.  We find that our calibrations at z = 5 for O3O2 and Ne3O2 do not reach values high enough to match the observed line ratios. Our model underestimates the line ratios by about 0.5 - 0.7 dex. Some of this mismatch might be explained further evolution of O3O2 and Ne3O2 towards higher values with increasing redshift beyond z = 5. As for R3, R2, and R32, we find that most of the observed galaxies lie either outside or at the edges of the 90\% confidence interval. These preliminary observations suggest that there may be some evolution of nebular metallicity indicators that are not captured by our models. 

The origin of this offset may be due to factors like varying ionization parameter, increasing hydrogen density with redshift, or increasing hardness of the ionizing spectrum  at fixed nebular oxygen abundance with redshift which is supported by studies like studies by \citet{strom2018}, \citet{Sanders_direct2019}, \citet{Topping2020a}, \citet{Topping2020b}, and \citet{Runco2021} as being the primary driver behind the evolution of nebular lines with redshift. We are limited in our modeling by the things we can control. Therefore we chose to manually vary the ionization parameter and see how it affects the line ratios. We find that  increasing ionization parameter can be one of the possible explanations for the observed redshift evolution of line ratios. This is supported by the fact that we systematically underproduce higher ionization lines like [O\scalebox{0.9}{ III}]$\lambda 5007$ and overproduce lower ionization lines like [O\scalebox{0.9}{ II}]$\lambda 3727, 29 $ implies that high redshift galaxies evolving towards having a higher ionization parameter as compared to their local counterparts. 

The ionization parameter encapsulates the ionization state of the nebula with a higher ionization parameter corresponding to more elements being in a higher ionization state. In our model, we define the ionization parameter as follows

\begin{equation}\label{u_eqn_hii}
U = {\Big(\frac{Q}{4\pi\,{R_s}^2\,n{_H}\,c}\Big)}    
\end{equation}

Where Q is the rate of ionizing photons in s$^{-1}$, n$_H$ is the hydrogen density in cm$^{-3}$ and R$_S$ is the Str\"{o}mgren sphere radius given by equation \ref{rs_eqn}. On substituting for the Str\"{o}mgren sphere radius defined in we get

\begin{equation}\label{u_prop}
U \propto Q^{1/3} n{_H}^{1/3}
\end{equation}

To test whether having a higher ionization parameter improves the agreement between our calibrations and the observations we have re-run our model with an increased hydrogen density of 100 cm$^{-3}$ and we set the inner radius of the cloud to be $10^{17}$ cm which changes the cloud geometry from a thick shell to spherical. As we move to a spherical geometry the cloud gets thicker and the inner radius moves closer to the star particle. This leads to an increase in higher ionization lines like [O\scalebox{0.9}{ III}]$\lambda 5007$ which is mainly produced in the inner regions of the cloud whereas the opposite is true for singly ionized lines like [O\scalebox{0.9}{ II}]$\lambda 3727, 29 $. In effect moving to a spherical geometry leads to an increase in ionization parameter. For more information on how the line ratios are affected by cloud geometry see \citet{garg2022bpt}. The resulting calibrations with a higher ionization parameter are shown as gray lines in Figure \ref{fig:jwst_mi_comp}. We tabulate the fits to the metallicity calibrations with a higher ionization parameter in Table \ref{tab:mi_highz} in (see Appendix \ref{app:mi_higZ}). 

As can be seen, having a higher ionization parameter leads to a much better match with the observations. This suggests that metallicity calibrations might evolve with redshift towards a higher ionization parameter. This is also supported by a number of studies \citep[e.g.][]{Brinchmann2004, liu2008, hainline2009, erb2010, wuyts2012, nakajima2013, shirazi2014, hayashi2015, onodera2016, Kashino_2017, reddy_2023a, reddy_2023b} that concluded that galaxies at $z \sim 2$ have higher ionization parameter compared to their local counterparts. Further, studies such as  \citet{curti_2022}, \citet{trump_2022} and, \citet{cameron_2023} have found that nebular line ratios might evolve with redshift since none of these metallicity indicators can simultaneously match the observations. \citet{trump_2022}, using photoionization modeling, concluded that the line ratios might be explained with an ISM with high ionization, low metallicity, and very high pressure suggesting extreme conditions in these high-$z$ galaxies. On the other hand \citet{nakajima_2023} found no strong redshift evolution for R2, R3, and R23. Whereas \citet{sanders_direct_2023} use a combined sample of 41 galaxies at redshifts ranging from 1.4 to 8.7 and find a clear evolution in high redshift metallicity indicators when compared to the z = 0 sample. However, similar to \citet{sanders_2023}, \citet{sanders_direct_2023} find no clear evolution amongst metallicity indicators from z = 2 onward.

Thus, the answer to the question of whether the nebular line ratios evolve with redshift is still unclear but preliminary observations do point to an evolution towards an increasing ionization parameter with redshift. The origin of this putative rise in a higher ionization parameter is unclear.  A non-exhaustive list includes increased hydrogen density in H\scalebox{0.9}{ II}regions \citet{reddy_2023a, reddy_2023b}, evolving cloud geometries, or an increased ionizing photon production rate (due to stellar processes such as binary star evolution, or stellar IMF variation).

On the theoretical end, \citet{hirschmann_2023} recently investigated the redshift evolution of strong line metallicity indicators in z $\sim$ 0 - 8 galaxies from the IllustrisTNG simulation. Their nebular emission model takes into account contributions from young stars, post-AGB stars, AGNs, and for the first time fast radiating shocks. This said, there are some significant differences between their methodology and the one employed in this study.  For example, \citet{hirschmann_2023} does not include dust radiative transfer and contribution from diffuse ionized gas whereas we do not include the contribution from radiating shocks and AGNs in our model. One major difference is that \citet{hirschmann_2023} model nebular emission from H\scalebox{0.9}{ II} region and Post-AGB stars by assigning each galaxy a single model that is selected based on the galaxy average properties like the metallicity, ionization parameter, and the C/O ratio using a pre-computed set of lookup tables. In contrast to this, we compute the nebular line emission on a particle-by-particle basis where each particle is assigned as an H\scalebox{0.9}{ II} region, post-AGB or diffuse ionized gas source which is modeled based on the properties of that particle. In the end, \citet{hirschmann_2023} conclude that strong line metallicity indicators evolve with redshift, and the evolution can be explained by high redshift galaxies having a higher ionization parameter which can be a by-product of then having a larger specific-SFR and higher assumed H\scalebox{0.9}{ II}region densities. 


\subsection{Caveats and Uncertainties}\label{sec:caveats}

\paragraph{N/O ratio}
\citet{garg2022bpt} found that the \textsc{simba} simulation is unable to reproduce the observed log(N/O) vs log(O/H) relation from \citet{pilyugin2012}, which may be attributed to the complexity involved in modeling nitrogen nucleosynthesis. The complexity arises because nitrogen can be produced via two pathways. In stars, nitrogen is produced through the CNO cycle where the reaction $ \mathrm{^{14}N(P,\gamma)^{15}O}$ is the slowest in the chain causing nitrogen to build up over time. If carbon synthesized within the star is used as a catalyst for this reaction, then the nitrogen produced is called primary nitrogen. Alternatively, if pre-existing carbon is used then it is called secondary nitrogen \citep{meynet2002origin}. 
 
In our models, we side-stepped the inability of {\sc simba} to match the local N/O vs O/H observed relationship by manually setting the nitrogen abundance as per equation \ref{pilyugin_eqn}. That said, if it turns out that the log(N/O) vs log(O/H) relation evolves with redshift then our model calibrations will not apply at high redshift. The possible evolution (or lack thereof) of N/O ratios with redshift is currently under debate. Some studies such as \citet{steidel2016} find no evolution with redshift whereas others find an increase in log(N/O) of anywhere between $0.1$ to $0.4$ dex at redshifts of about $1.5$ \citep{masters2014, strom2018, bian2020}.

\paragraph{Sub-resolution effects}
Current state-of-the-art large-scale cosmological galaxy formation simulations (including \textsc{simba}) are unable to achieve a high enough resolution to directly model individual H\scalebox{0.9}{ II} regions or diffuse ionized gas clouds. Due to this, we have to resort to sub-grid modeling which introduces uncertainties to our model. We enforce $4$ major assumptions that could substantively impact our model results:
\begin{enumerate}
    \item{\it Mass and age distribution:} A single particle in our \textsc{simba} box is of the order of $2.3\times10^6$ M$\odot$ which is too massive to be modeled as a single H\scalebox{0.9}{ II} region. Further \textsc{simba} has a stochastic implementation of stellar ages for young star particles. To counter these drawbacks we divide each star particle into smaller mass particles with varying ages based on power-law distributions. We have kept the slope of this distribution fixed across redshifts and different physical conditions which introduces a level of uncertainty in our modeling. 
    
    \item {\it Geometry:} Another major sub-resolution model is in our assumption of spherical H\scalebox{0.9}{ II} regions: we are unable to model the true geometry and escape fractions of H\scalebox{0.9}{ II} regions surrounding massive stars. Similarly, the detailed geometry of the DIG is unresolved, and we are instead forced to assume a plane parallel geometry. 
    
    \item {\it Density:} We assume a fixed density for H\scalebox{0.9}{ II} regions and DIG gas that does not vary with redshift. Increasing hydrogen density will lead to an increase in the average ionization parameter thus the flux of singly ionized species like [N\scalebox{0.9}{ II}]$\lambda 6583$, [S\scalebox{0.9}{ II}]$\lambda 6717, 31$ will decrease whereas the flux of doubly ionized species like [O\scalebox{0.9}{ III}]$\lambda 5007$ will increase.
\end{enumerate}

\paragraph{Other sources of ionizing radiation}
Though we have expanded our base model from \citet{garg2022bpt} to include nebular emission not only from H\scalebox{0.9}{ II} regions but also post-AGB stars and diffused ionized gas we are still missing some other sources of ionizing radiation in a galaxy namely active galactic nuclei (AGNs) and fast radiating shocks. These can be a substantial source of ionizing radiation and including them might be important to accurately model nebular emission across redshifts.

\paragraph{Diffuse Ionized Gas Modeling} The main uncertainty in our diffused ionized gas model comes from our approximation of the shape of the input radiation field impinging on the gas cell. As described in detail in section \secref{sec:method_sources} we compute the shape of the radiation field by taking a distance weighted average of the \textsc{cloudy} output spectrum of the ionizing radiation sources (H\scalebox{0.9}{ II} regions around massive stars and post-AGB stars) within 1 kpc. This implicitly assumes the optically thin transfer of those photons to the incident DIG cell. We additionally are required to make assumptions regarding the escape fractions from H\scalebox{0.9}{ II} regions and post-AGB stars which as described in \secref{sec:method_sources} are kept fixed at $0.4$ and $0.6$ respectively. This, of course, does not capture the wide variety of environments that these ionizing sources can be found in. 

Beyond this, our mass resolution in cosmological simulations prevents us from modeling directly the geometry of DIG gas cells. The \textsc{simba} snapshots used in this study we have a resolution of about 10$^6$ M$_{\odot}$ which means that we can model only a collection of gas clouds at a time that fails to capture any potential substructure. In reality, individual gas clouds span a wide range of densities and geometry.

Taken together, these uncertainties in our DIG modeling may explain why our calibrations for \textsc{simba} z = 0 galaxies involving [O\scalebox{0.9}{ III}]$\lambda 5007$, [O\scalebox{0.9}{ II}]$\lambda 3727, 29 $ and [S\scalebox{0.9}{ II}]$\lambda 6717, 31$ like R2, R3, R23, O3O2, and S2 match the auroral line observations fairly well at low metallicities but are not as good a match at high metallicities (see Figure \ref{fig:z0_all} and section \ref{sec:local_mi} for more detail). This coupled with the fact that our models are a reasonably good match to N2 which does not have a substantial DIG contribution at high metallicities strongly suggests that the mismatch might be explained by the inability of our model to accurately capture the diffused ionized gas emission. 

\section{Summary}\label{sec:summary}
In summary, in this study, we have investigated the redshift evolution of 12 strong line metallicity indicators. We modeled nebular emission on a particle-by-particle basis on galaxies from the \textsc{simba} using \textsc{cloudy} photoionization code. We had expanded our model to not only include the contribution from H\scalebox{0.9}{ II} regions but also post-AGB stars and diffuse ionized gas. The main results of this study are as follows:

\begin{itemize}
\item We find that our model reasonably matches the observed auroral line calibrations and reproduces the general nature of the curve for most of the line ratios (\secref{sec:local_mi}, Figure \ref{fig:z0_all}, Table \ref{tab:mi_z0}).

\item We run our models on the \textsc{simba} cosmological simulation at integer redshifts from $z=0-5$, and find mild redshift evolution for all the 12 line ratios (\secref{sec:z_mi}, Figure \ref{fig:z_all}). This suggests that are affected by properties that vary in our model with redshift, which include: the spectrum of ionizing photons in H\scalebox{0.9}{ II} regions, and post-AGB stars, incident radiation field for calculating DIG emission, the stellar IMF, and the amount of gas available for ionization have a discernible impact on the metallicity indicators. 

\item We use these results to assess strong line calibrations at high-redshift (\secref{sec:z_mi}, Figure \ref{fig:z_all}, Appendix \ref{app:mi_higZ}). We find that ratios R3, R2, and R32 show no evolution with redshift. Whereas other line ratios like N2, O3N2, R3N2, N2O2, O3O2, Ne3O2, S2, O3S2 and N2S2 show a an evolution raging from 0.2 - 0.5 dex between z = 0 and z = 5. Further, we find thatR3, R2, and R32 work fairly well at low metallicities while N2, O3N2, R3N2, N2O2, O3S2, and N2S2 work well at high metallicities / low redshifts. On the other hand, we find O3O2 and Ne3O2 have limited applicability owing to their flat slopes and fairly wide confidence intervals.

\item We find that for local galaxies diffuse ionized gas contribution increases with metallicity for singly ionized oxygen and sulfur lines. It can account for as much as 70\% of the total line emission in extreme cases (Figure \ref{fig:rc}). This demonstrates the importance of including the contribution from DIG when modeling nebular emission. 

\item On comparing our metallicity calibrations against recent JWST auroral line measurements we see a slight mismatch and find that having a higher ionization parameter at high redshifts might be one of the possible explanations (Figure \ref{fig:jwst_mi_comp}). The increased ionization parameter could be caused by increasing hydrogen density, evolving cloud geometry, or an evolving (more top-heavy) IMF with redshift. Current observations are limited to a handful of galaxies and more JWST observations in the coming years will enable us to conclusively answer whether the metallicity indicators evolve with redshift or not.  

\end{itemize}

\section{Acknowledgements}
The authors would like to thank Allison Strom for fruitful discussions that guided the analysis of this paper. Some of the \textsc{simba} simulations boxes used in this work were run on the Hipergator computing cluster at the University of Florida. The flagship \textsc{simba} simulation was run using the DiRAC@Durham facility managed by the Institute for Computational Cosmology on behalf of the STFC DiRAC HPC Facility. The equipment was funded by BEIS capital funding via STFC capital grants ST/P002293/1, ST/R002371/1, and ST/S002502/1, Durham University, and STFC operations grant ST/R000832/1. R.D. acknowledges support from the Wolfson Research Merit Award program of the U.K. Royal Society. P.G. and D.N. were funded by JWST-AR-01883.001.

\software{caesar \citep{Thompson2014}, cloudy \citep[v17.00][]{ferland2017}, cloudyfsps \citep{byler2017}, fsps \citep{conroy2009,conroy2010}, powderday \citep{powderday}, python-fsps \citep{python_fsps}, simba \citep{simba}}

\bibliographystyle{aasjournal}
\bibliography{references}

\begin{thebibliography}{}
\expandafter\ifx\csname natexlab\endcsname\relax\def\natexlab#1{#1}\fi
\providecommand{\url}[1]{\href{#1}{#1}}
\providecommand{\dodoi}[1]{doi:~\href{http://doi.org/#1}{\nolinkurl{#1}}}
\providecommand{\doeprint}[1]{\href{http://ascl.net/#1}{\nolinkurl{http://ascl.net/#1}}}
\providecommand{\doarXiv}[1]{\href{https://arxiv.org/abs/#1}{\nolinkurl{https://arxiv.org/abs/#1}}}

\bibitem[{Andrews \& Martini(2013)}]{andrews2013}
Andrews, B.~H., \& Martini, P. 2013, The Astrophysical Journal, 765, 140, \dodoi{10.1088/0004-637x/765/2/140}

\bibitem[{Angl{\'e}s-Alc{\'a}zar {et~al.}(2017{\natexlab{a}})Angl{\'e}s-Alc{\'a}zar, Dav{\'e}, Faucher-Gigu{\`e}re, {\"O}zel, \& Hopkins}]{angles2017a}
Angl{\'e}s-Alc{\'a}zar, D., Dav{\'e}, R., Faucher-Gigu{\`e}re, C.-A., {\"O}zel, F., \& Hopkins, P.~F. 2017{\natexlab{a}}, Monthly Notices of the Royal Astronomical Society, 464, 2840

\bibitem[{Angl{\'e}s-Alc{\'a}zar {et~al.}(2017{\natexlab{b}})Angl{\'e}s-Alc{\'a}zar, Faucher-Gigu{\`e}re, Kere{\v{s}}, Hopkins, Quataert, \& Murray}]{angles2017b}
Angl{\'e}s-Alc{\'a}zar, D., Faucher-Gigu{\`e}re, C.-A., Kere{\v{s}}, D., {et~al.} 2017{\natexlab{b}}, Monthly Notices of the Royal Astronomical Society, 470, 4698

\bibitem[{Arellano-Córdova {et~al.}(2022)Arellano-Córdova, Berg, Chisholm, Haro, Dickinson, Finkelstein, Leclercq, Rogers, Simons, Skillman, Trump, \& Kartaltepe}]{arellano-cordova_2022}
Arellano-Córdova, K.~Z., Berg, D.~A., Chisholm, J., {et~al.} 2022, The Astrophysical Journal Letters, 940, L23, \dodoi{10.3847/2041-8213/ac9ab2}

\bibitem[{Bayliss {et~al.}(2014)Bayliss, Rigby, Sharon, Wuyts, Florian, Gladders, Johnson, \& Oguri}]{Bayliss_2014}
Bayliss, M.~B., Rigby, J.~R., Sharon, K., {et~al.} 2014, The Astrophysical Journal, 790, 144, \dodoi{10.1088/0004-637x/790/2/144}

\bibitem[{Berg {et~al.}(2018)Berg, Erb, Auger, Pettini, \& Brammer}]{Berg_2018}
Berg, D.~A., Erb, D.~K., Auger, M.~W., Pettini, M., \& Brammer, G.~B. 2018, The Astrophysical Journal, 859, 164, \dodoi{10.3847/1538-4357/aab7fa}

\bibitem[{Berg {et~al.}(2015)Berg, Skillman, Croxall, Pogge, Moustakas, \& Johnson-Groh}]{Berg_2015}
Berg, D.~A., Skillman, E.~D., Croxall, K.~V., {et~al.} 2015, The Astrophysical Journal, 806, 16, \dodoi{10.1088/0004-637x/806/1/16}

\bibitem[{Bian {et~al.}(2020)Bian, Kewley, Groves, \& Dopita}]{bian2020}
Bian, F., Kewley, L.~J., Groves, B., \& Dopita, M.~A. 2020, Monthly Notices of the Royal Astronomical Society, 493, 580

\bibitem[{Bondi \& Hoyle(1944)}]{bondi1944}
Bondi, H., \& Hoyle, F. 1944, Monthly Notices of the Royal Astronomical Society, 104, 273

\bibitem[{Bowen(1927)}]{bowen1927}
Bowen, I.~S. 1927, Nature, 120, 473

\bibitem[{Brammer {et~al.}(2012)Brammer, S{\'{a}}nchez-Janssen, Labb{\'{e}}, da~Cunha, Erb, Franx, Fumagalli, Lundgren, Marchesini, Momcheva, Nelson, Patel, Quadri, Rix, Skelton, Schmidt, van~der Wel, van Dokkum, Wake, \& Whitaker}]{Brammer_2012}
Brammer, G.~B., S{\'{a}}nchez-Janssen, R., Labb{\'{e}}, I., {et~al.} 2012, The Astrophysical Journal, 758, L17, \dodoi{10.1088/2041-8205/758/1/l17}

\bibitem[{Bresolin(2007)}]{Bresolin_2007}
Bresolin, F. 2007, The Astrophysical Journal, 656, 186, \dodoi{10.1086/510380}

\bibitem[{Brinchmann(2022)}]{brinchmann_2022}
Brinchmann, J. 2022, High-z galaxies with {JWST} and local analogues -- it is not only star formation,  arXiv, \dodoi{10.48550/arXiv.2208.07467}

\bibitem[{Brinchmann {et~al.}(2004)Brinchmann, Charlot, White, Tremonti, Kauffmann, Heckman, \& Brinkmann}]{Brinchmann2004}
Brinchmann, J., Charlot, S., White, S. D.~M., {et~al.} 2004, Monthly Notices of the Royal Astronomical Society, 351, 1151, \dodoi{10.1111/j.1365-2966.2004.07881.x}

\bibitem[{Brown {et~al.}(2016)Brown, Martini, \& Andrews}]{brown_2016}
Brown, J.~S., Martini, P., \& Andrews, B.~H. 2016, Monthly Notices of the Royal Astronomical Society, 458, 1529, \dodoi{10.1093/mnras/stw392}

\bibitem[{Byler {et~al.}(2017)Byler, Dalcanton, Conroy, \& Johnson}]{byler2017}
Byler, N., Dalcanton, J.~J., Conroy, C., \& Johnson, B.~D. 2017, The Astrophysical Journal, 840, 44

\bibitem[{Byler {et~al.}(2019)Byler, Dalcanton, Conroy, Johnson, Choi, Dotter, \& Rosenfield}]{byler_2019}
Byler, N., Dalcanton, J.~J., Conroy, C., {et~al.} 2019, The Astronomical Journal, 158, 2, \dodoi{10.3847/1538-3881/ab1b70}

\bibitem[{Cameron {et~al.}(2023)Cameron, Saxena, Bunker, D'Eugenio, Carniani, Maiolino, Curtis-Lake, Ferruit, Jakobsen, Arribas, Bonaventura, Charlot, Chevallard, Curti, Looser, Maseda, Rawle, Del~Pino, Smit, Übler, Willott, Witstok, Egami, Eisenstein, Johnson, Hainline, Rieke, Robertson, Stark, Tacchella, Williams, Bhatawdekar, Bowler, Boyett, Circosta, Helton, Jones, Kumari, Ji, Nelson, Parlanti, Sandles, Scholtz, \& Sun}]{cameron_2023}
Cameron, A.~J., Saxena, A., Bunker, A.~J., {et~al.} 2023, \dodoi{10.48550/arxiv.2302.04298}

\bibitem[{Chandar {et~al.}(2014)Chandar, Whitmore, Calzetti, \& O'Connell}]{chandar2014}
Chandar, R., Whitmore, B.~C., Calzetti, D., \& O'Connell, R. 2014, The Astrophysical Journal, 787, 17

\bibitem[{Chandar {et~al.}(2016)Chandar, Whitmore, Dinino, Kennicutt, Chien, Schinnerer, \& Meidt}]{chandar2016}
Chandar, R., Whitmore, B.~C., Dinino, D., {et~al.} 2016, The Astrophysical Journal, 824, 71

\bibitem[{Choi {et~al.}(2020)Choi, Dalcanton, Williams, Skillman, Fouesneau, Gordon, Sandstrom, Weisz, \& Gilbert}]{choi_2020}
Choi, Y., Dalcanton, J.~J., Williams, B.~F., {et~al.} 2020, The Astrophysical Journal, 902, 54, \dodoi{10.3847/1538-4357/abb467}

\bibitem[{Christensen {et~al.}(2012{\natexlab{a}})Christensen, Richard, Hjorth, Milvang-Jensen, Laursen, Limousin, Dessauges-Zavadsky, Grillo, \& Ebeling}]{Christensen2012a}
Christensen, L., Richard, J., Hjorth, J., {et~al.} 2012{\natexlab{a}}, Monthly Notices of the Royal Astronomical Society, 427, 1953, \dodoi{10.1111/j.1365-2966.2012.22006.x}

\bibitem[{Christensen {et~al.}(2012{\natexlab{b}})Christensen, Laursen, Richard, Hjorth, Milvang-Jensen, Dessauges-Zavadsky, Limousin, Grillo, \& Ebeling}]{Christensen2012b}
Christensen, L., Laursen, P., Richard, J., {et~al.} 2012{\natexlab{b}}, Monthly Notices of the Royal Astronomical Society, 427, 1973, \dodoi{10.1111/j.1365-2966.2012.22007.x}

\bibitem[{Conroy \& Gunn(2010)}]{conroy2010}
Conroy, C., \& Gunn, J.~E. 2010, The Astrophysical Journal, 712, 833

\bibitem[{Conroy {et~al.}(2009)Conroy, Gunn, \& White}]{conroy2009}
Conroy, C., Gunn, J.~E., \& White, M. 2009, The Astrophysical Journal, 699, 486

\bibitem[{Croxall {et~al.}(2015)Croxall, Pogge, Berg, Skillman, \& Moustakas}]{Croxall_2015}
Croxall, K.~V., Pogge, R.~W., Berg, D.~A., Skillman, E.~D., \& Moustakas, J. 2015, The Astrophysical Journal, 808, 42, \dodoi{10.1088/0004-637x/808/1/42}

\bibitem[{Croxall {et~al.}(2016)Croxall, Pogge, Berg, Skillman, \& Moustakas}]{Croxall_2016}
---. 2016, The Astrophysical Journal, 830, 4, \dodoi{10.3847/0004-637x/830/1/4}

\bibitem[{Curti {et~al.}(2017)Curti, Cresci, Mannucci, Marconi, Maiolino, \& Esposito}]{curti_2017}
Curti, M., Cresci, G., Mannucci, F., {et~al.} 2017, Monthly Notices of the Royal Astronomical Society, 465, 1384, \dodoi{10.1093/mnras/stw2766}

\bibitem[{Curti {et~al.}(2020)Curti, Mannucci, Cresci, \& Maiolino}]{curti_2020}
Curti, M., Mannucci, F., Cresci, G., \& Maiolino, R. 2020, Monthly Notices of the Royal Astronomical Society, 491, 944, \dodoi{10.1093/mnras/stz2910}

\bibitem[{Curti {et~al.}(2022)Curti, D’Eugenio, Carniani, Maiolino, Sandles, Witstok, Baker, Bennett, Piotrowska, Tacchella, Charlot, Nakajima, Maheson, Mannucci, Amiri, Arribas, Belfiore, Bonaventura, Bunker, Chevallard, Cresci, Curtis-Lake, Hayden-Pawson, Jones, Kumari, Laseter, Looser, Marconi, Maseda, Scholtz, Smit, Übler, \& Wallace}]{curti_2022}
Curti, M., D’Eugenio, F., Carniani, S., {et~al.} 2022, Monthly Notices of the Royal Astronomical Society, 518, 425, \dodoi{10.1093/mnras/stac2737}

\bibitem[{Dav{\'e} {et~al.}(2019)Dav{\'e}, Angl{\'e}s-Alc{\'a}zar, Narayanan, Li, Rafieferantsoa, \& Appleby}]{simba}
Dav{\'e}, R., Angl{\'e}s-Alc{\'a}zar, D., Narayanan, D., {et~al.} 2019, Monthly Notices of the Royal Astronomical Society, 486, 2827

\bibitem[{Dav{\'e} {et~al.}(2016)Dav{\'e}, Thompson, \& Hopkins}]{dave2016mufasa}
Dav{\'e}, R., Thompson, R., \& Hopkins, P.~F. 2016, Monthly Notices of the Royal Astronomical Society, 462, 3265

\bibitem[{Della~Bruna {et~al.}(2021)Della~Bruna, Adamo, Lee, Smith, Krumholz, Bik, Calzetti, Fox, Fumagalli, Grasha, Messa, Östlin, Walterbos, \& Wofford}]{della_bruna_2021}
Della~Bruna, L., Adamo, A., Lee, J.~C., {et~al.} 2021, Astronomy \& Astrophysics, 650, A103, \dodoi{10.1051/0004-6361/202039402}

\bibitem[{Denicoló {et~al.}(2002)Denicoló, Terlevich, \& Terlevich}]{denicolo_2002}
Denicoló, G., Terlevich, R., \& Terlevich, E. 2002, Monthly Notices of the Royal Astronomical Society, 330, 69, \dodoi{10.1046/j.1365-8711.2002.05041.x}

\bibitem[{Dopita {et~al.}(2016)Dopita, Kewley, Sutherland, \& Nicholls}]{dopita2016}
Dopita, M.~A., Kewley, L.~J., Sutherland, R.~S., \& Nicholls, D.~C. 2016, Astrophysics and Space Science, 361, 61

\bibitem[{Eldridge {et~al.}(2017)Eldridge, Stanway, Xiao, McClelland, Taylor, Ng, Greis, \& Bray}]{eldridge2017}
Eldridge, J., Stanway, E., Xiao, L., {et~al.} 2017, Publications of the Astronomical Society of Australia, 34

\bibitem[{Erb {et~al.}(2010)Erb, Pettini, Shapley, Steidel, Law, \& Reddy}]{erb2010}
Erb, D.~K., Pettini, M., Shapley, A.~E., {et~al.} 2010, The Astrophysical Journal, 719, 1168

\bibitem[{{Ferland} {et~al.}(2017){Ferland}, {Chatzikos}, {Guzm{\'a}n}, {Lykins}, {van Hoof}, {Williams}, {Abel}, {Badnell}, {Keenan}, {Porter}, \& {Stancil}}]{ferland2017}
{Ferland}, G.~J., {Chatzikos}, M., {Guzm{\'a}n}, F., {et~al.} 2017, \rmxaa, 53, 385.
\newblock \doarXiv{1705.10877}

\bibitem[{Garg {et~al.}(2022)Garg, Narayanan, Byler, Sanders, Shapley, Strom, Davé, Hirschmann, Lovell, Otter, Popping, \& Privon}]{garg2022bpt}
Garg, P., Narayanan, D., Byler, N., {et~al.} 2022, The Astrophysical Journal, 926, 80, \dodoi{10.3847/1538-4357/ac43b8}

\bibitem[{Gburek {et~al.}(2019)Gburek, Siana, Alavi, Emami, Richard, Freeman, Stark, Snapp-Kolas, \& Lucero}]{Gburek_2019}
Gburek, T., Siana, B., Alavi, A., {et~al.} 2019, The Astrophysical Journal, 887, 168, \dodoi{10.3847/1538-4357/ab5713}

\bibitem[{Haardt \& Madau(2012)}]{haardt_2012}
Haardt, F., \& Madau, P. 2012, The Astrophysical Journal, 746, 125, \dodoi{10.1088/0004-637X/746/2/125}

\bibitem[{Hainline {et~al.}(2009)Hainline, Shapley, Kornei, Pettini, Buckley-Geer, Allam, \& Tucker}]{hainline2009}
Hainline, K.~N., Shapley, A.~E., Kornei, K.~A., {et~al.} 2009, The Astrophysical Journal, 701, 52, \dodoi{10.1088/0004-637x/701/1/52}

\bibitem[{Hayashi {et~al.}(2015)Hayashi, Ly, Shimasaku, Motohara, Malkan, Nagao, Kashikawa, Goto, \& Naito}]{hayashi2015}
Hayashi, M., Ly, C., Shimasaku, K., {et~al.} 2015, Publications of the Astronomical Society of Japan, 67, \dodoi{10.1093/pasj/psv041}

\bibitem[{Heintz {et~al.}(2023)Heintz, Giménez-Arteaga, Fujimoto, Brammer, Espada, Gillman, González-López, Greve, Harikane, Hatsukade, Knudsen, Koekemoer, Kohno, Kokorev, Lee, Magdis, Nelson, Rizzo, Sanders, Schaerer, Shapley, Strait, Toft, Valentino, van~der Wel, Vijayan, Watson, Bauer, Christiansen, \& Wilson}]{heintz_2023}
Heintz, K.~E., Giménez-Arteaga, C., Fujimoto, S., {et~al.} 2023, The gas and stellar content of a metal-poor galaxy at \$z=8.496\$ as revealed by {JWST} and {ALMA},  arXiv, \dodoi{10.48550/arXiv.2212.06877}

\bibitem[{Henry {et~al.}(2018)Henry, Stephenson, Miller~Bertolami, Kwitter, \& Balick}]{henry_2018}
Henry, R. B.~C., Stephenson, B.~G., Miller~Bertolami, M.~M., Kwitter, K.~B., \& Balick, B. 2018, Monthly Notices of the Royal Astronomical Society, 473, 241, \dodoi{10.1093/mnras/stx2286}

\bibitem[{Hirschmann {et~al.}(2023)Hirschmann, Charlot, \& Somerville}]{hirschmann_2023}
Hirschmann, M., Charlot, S., \& Somerville, R.~S. 2023, High-redshift metallicity calibrations for {JWST} spectra: insights from line emission in cosmological simulations,  arXiv, \dodoi{10.48550/arXiv.2305.03753}

\bibitem[{Hough {et~al.}(2023)Hough, Rennehan, Kobayashi, Loubser, Davé, Babul, \& Cui}]{hough_simba-c_2023}
Hough, R.~T., Rennehan, D., Kobayashi, C., {et~al.} 2023, Monthly Notices of the Royal Astronomical Society, stad2394, \dodoi{10.1093/mnras/stad2394}

\bibitem[{Izotov {et~al.}(2006)Izotov, Stasi\'{}nska, Meynet, Guseva, \& Thuan}]{Izotov2006}
Izotov, Y.~I., Stasi\'{}nska, G., Meynet, G., Guseva, N.~G., \& Thuan, T.~X. 2006, A\&A, 448, 955, \dodoi{10.1051/0004-6361:20053763}

\bibitem[{James {et~al.}(2014)James, Pettini, Christensen, Auger, Becker, King, Quider, Shapley, \& Steidel}]{James2014}
James, B.~L., Pettini, M., Christensen, L., {et~al.} 2014, Monthly Notices of the Royal Astronomical Society, 440, 1794, \dodoi{10.1093/mnras/stu287}

\bibitem[{Johnson {et~al.}(2021)Johnson, Foreman-Mackey, Sick, Leja, Byler, Walmsley, Tollerud, Leung, \& Scott}]{python_fsps}
Johnson, B., Foreman-Mackey, D., Sick, J., {et~al.} 2021, dfm/python-fsps: python-fsps v0.4.1rc1, v0.4.1rc1,  Zenodo, \dodoi{10.5281/zenodo.4737461}

\bibitem[{Jones {et~al.}(2015)Jones, Martin, \& Cooper}]{Jones2015}
Jones, T., Martin, C., \& Cooper, M.~C. 2015, The Astrophysical Journal, 813, 126, \dodoi{10.1088/0004-637x/813/2/126}

\bibitem[{Kashino {et~al.}(2017)Kashino, Silverman, Sanders, Kartaltepe, Daddi, Renzini, Valentino, Rodighiero, Juneau, Kewley, Zahid, Arimoto, Nagao, Chu, Sugiyama, Civano, Ilbert, Kajisawa, F{\`{e}}vre, Maier, Masters, Miyaji, Onodera, Puglisi, \& Taniguchi}]{Kashino_2017}
Kashino, D., Silverman, J.~D., Sanders, D., {et~al.} 2017, The Astrophysical Journal, 835, 88, \dodoi{10.3847/1538-4357/835/1/88}

\bibitem[{Katz {et~al.}(2022)Katz, Saxena, Cameron, Carniani, Bunker, Arribas, Bhatawdekar, Bowler, Boyett, Cresci, Curtis-Lake, D’Eugenio, Kumari, Looser, Maiolino, Übler, Willott, \& Witstok}]{katz_2022}
Katz, H., Saxena, A., Cameron, A.~J., {et~al.} 2022, Monthly Notices of the Royal Astronomical Society, 518, 592, \dodoi{10.1093/mnras/stac2657}

\bibitem[{Keenan {et~al.}(1996)Keenan, Aller, Bell, Hyung, McKenna, \& Ramsbottom}]{Keenan1996}
Keenan, F.~P., Aller, L.~H., Bell, K.~L., {et~al.} 1996, Monthly Notices of the Royal Astronomical Society, 281, 1073, \dodoi{10.1093/mnras/281.3.1073}

\bibitem[{Kennicutt(1998)}]{kennicutt_1998}
Kennicutt, Jr., R.~C. 1998, The Astrophysical Journal, 498, 541, \dodoi{10.1086/305588}

\bibitem[{Kewley \& Dopita(2002)}]{kewley2002}
Kewley, L.~J., \& Dopita, M.~A. 2002, The Astrophysical Journal Supplement Series, 142, 35

\bibitem[{Kojima {et~al.}(2017)Kojima, Ouchi, Nakajima, Shibuya, Harikane, \& Ono}]{Kojima2017}
Kojima, T., Ouchi, M., Nakajima, K., {et~al.} 2017, Publications of the Astronomical Society of Japan, 69, \dodoi{10.1093/pasj/psx017}

\bibitem[{Krumholz {et~al.}(2009)Krumholz, McKee, \& Tumlinson}]{krumholz2009}
Krumholz, M.~R., McKee, C.~F., \& Tumlinson, J. 2009, The Astrophysical Journal, 693, 216

\bibitem[{Langeroodi {et~al.}(2022)Langeroodi, Hjorth, Chen, Kelly, Williams, Lin, Scarlata, Zitrin, Broadhurst, Diego, Huang, Filippenko, Foley, Jha, Koekemoer, Oguri, Perez-Fournon, Pierel, Poidevin, \& Strolger}]{langeroodi_2022}
Langeroodi, D., Hjorth, J., Chen, W., {et~al.} 2022, Evolution of the {Mass}-{Metallicity} {Relation} from {Redshift} \$z{\textbackslash}approx8\$ to the {Local} {Universe},  arXiv, \dodoi{10.48550/arXiv.2212.02491}

\bibitem[{Larson {et~al.}(2020)Larson, D{\'\i}az-Santos, Armus, Privon, Linden, Evans, Howell, Charmandaris, Sanders, Stierwalt, {et~al.}}]{larson2020}
Larson, K., D{\'\i}az-Santos, T., Armus, L., {et~al.} 2020, The Astrophysical Journal, 888, 92

\bibitem[{{Li} {et~al.}(2019){Li}, {Narayanan}, \& {Dav{\'e}}}]{li2019}
{Li}, Q., {Narayanan}, D., \& {Dav{\'e}}, R. 2019, \mnras, 490, 1425, \dodoi{10.1093/mnras/stz2684}

\bibitem[{Lin {et~al.}(2017)Lin, Hu, Kong, Gao, Zou, Wang, Cheng, Fang, Lin, \& Wang}]{Lin_2017}
Lin, Z., Hu, N., Kong, X., {et~al.} 2017, The Astrophysical Journal, 842, 97, \dodoi{10.3847/1538-4357/aa6f14}

\bibitem[{Linden {et~al.}(2017)Linden, Evans, Rich, Larson, Armus, D{\'\i}az-Santos, Privon, Howell, Inami, Kim, {et~al.}}]{linden2017}
Linden, S., Evans, A., Rich, J., {et~al.} 2017, The Astrophysical Journal, 843, 91

\bibitem[{Liu {et~al.}(2008)Liu, Shapley, Coil, Brinchmann, \& Ma}]{liu2008}
Liu, X., Shapley, A.~E., Coil, A.~L., Brinchmann, J., \& Ma, C.-P. 2008, The Astrophysical Journal, 678, 758

\bibitem[{Lovell {et~al.}(2021)Lovell, Geach, Davé, Narayanan, \& Li}]{lovell_2021}
Lovell, C.~C., Geach, J.~E., Davé, R., Narayanan, D., \& Li, Q. 2021, Monthly Notices of the Royal Astronomical Society, 502, 772, \dodoi{10.1093/mnras/staa4043}

\bibitem[{Maiolino {et~al.}(2008)Maiolino, Nagao, Grazian, Cocchia, Marconi, Mannucci, Cimatti, Pipino, Ballero, Calura, Chiappini, Fontana, Granato, Matteucci, Pastorini, Pentericci, Risaliti, Salvati, \& Silva}]{Maiolino2008}
Maiolino, R., Nagao, T., Grazian, A., {et~al.} 2008, A\&A, 488, 463, \dodoi{10.1051/0004-6361:200809678}

\bibitem[{Marino {et~al.}(2013)Marino, Rosales-Ortega, Sánchez, Gil~de Paz, Vílchez, Miralles-Caballero, Kehrig, Pérez-Montero, Stanishev, Iglesias-Páramo, Díaz, Castillo-Morales, Kennicutt, López-Sánchez, Galbany, García-Benito, Mast, Mendez-Abreu, Monreal-Ibero, Husemann, Walcher, García-Lorenzo, Masegosa, del Olmo~Orozco, Mourão, Ziegler, Mollá, Papaderos, Sánchez-Blázquez, González~Delgado, Falcón-Barroso, Roth, van~de Ven, \& team}]{marino_2013}
Marino, R.~A., Rosales-Ortega, F.~F., Sánchez, S.~F., {et~al.} 2013, Astronomy \& Astrophysics, 559, A114, \dodoi{10.1051/0004-6361/201321956}

\bibitem[{Maseda {et~al.}(2014)Maseda, van~der Wel, Rix, da~Cunha, Pacifici, Momcheva, Brammer, Meidt, Franx, van Dokkum, Fumagalli, Bell, Ferguson, Förster-Schreiber, Koekemoer, Koo, Lundgren, Marchesini, Nelson, Patel, Skelton, Straughn, Trump, \& Whitaker}]{Maseda_2014}
Maseda, M.~V., van~der Wel, A., Rix, H.-W., {et~al.} 2014, The Astrophysical Journal, 791, 17, \dodoi{10.1088/0004-637x/791/1/17}

\bibitem[{Masters {et~al.}(2014)Masters, McCarthy, Siana, Malkan, Mobasher, Atek, Henry, Martin, Rafelski, Hathi, {et~al.}}]{masters2014}
Masters, D., McCarthy, P., Siana, B., {et~al.} 2014, The Astrophysical Journal, 785, 153

\bibitem[{Matthee {et~al.}(2022)Matthee, Mackenzie, Simcoe, Kashino, Lilly, Bordoloi, \& Eilers}]{matthee_2022}
Matthee, J., Mackenzie, R., Simcoe, R.~A., {et~al.} 2022, {EIGER} {II}. first spectroscopic characterisation of the young stars and ionised gas associated with strong {H}\${\textbackslash}beta\$ and [{OIII}] line-emission in galaxies at z=5-7 with {JWST},  arXiv, \dodoi{10.48550/arXiv.2211.08255}

\bibitem[{Mc~Leod {et~al.}(2015)Mc~Leod, Weilbacher, Ginsburg, Dale, Ramsay, \& Testi}]{McLeod2015}
Mc~Leod, A.~F., Weilbacher, P.~M., Ginsburg, A., {et~al.} 2015, Monthly Notices of the Royal Astronomical Society, 455, 4057, \dodoi{10.1093/mnras/stv2617}

\bibitem[{McCall {et~al.}(1985)McCall, Rybski, \& Shields}]{mccall1985}
McCall, M.~L., Rybski, P., \& Shields, G. 1985, The Astrophysical Journal Supplement Series, 57, 1

\bibitem[{McGaugh(1991)}]{mcgaugh1991}
McGaugh, S.~S. 1991, The Astrophysical Journal, 380, 140

\bibitem[{McLennan \& Shrum(1925)}]{McLennan1925}
McLennan, J.~C., \& Shrum, G.~M. 1925, Proceedings of the Royal Society of London. Series A, Containing Papers of a Mathematical and Physical Character, 108, 501.
\newblock \url{http://www.jstor.org/stable/94431}

\bibitem[{Meynet \& Maeder(2002)}]{meynet2002origin}
Meynet, G., \& Maeder, A. 2002, Astronomy \& Astrophysics, 381, L25

\bibitem[{Monreal-Ibero {et~al.}(2012)Monreal-Ibero, Walsh, \& V\'{\i}lchez}]{Monreal2012}
Monreal-Ibero, A., Walsh, J.~R., \& V\'{\i}lchez, J.~M. 2012, A\&A, 544, A60, \dodoi{10.1051/0004-6361/201219543}

\bibitem[{Nagao {et~al.}(2006)Nagao, Maiolino, \& Marconi}]{nagao_2006}
Nagao, T., Maiolino, R., \& Marconi, A. 2006, Astronomy \& Astrophysics, 459, 85, \dodoi{10.1051/0004-6361:20065216}

\bibitem[{Nakajima {et~al.}(2023)Nakajima, Ouchi, Isobe, Harikane, Zhang, Ono, Umeda, \& Oguri}]{nakajima_2023}
Nakajima, K., Ouchi, M., Isobe, Y., {et~al.} 2023, {JWST} {Census} for the {Mass}-{Metallicity} {Star}-{Formation} {Relations} at z=4-10 with {Self}-{Consistent} {Flux} {Calibration} and the {Proper} {Metallicity} {Calibrators},  arXiv, \dodoi{10.48550/arXiv.2301.12825}

\bibitem[{Nakajima {et~al.}(2013)Nakajima, Ouchi, Shimasaku, Hashimoto, Ono, \& Lee}]{nakajima2013}
Nakajima, K., Ouchi, M., Shimasaku, K., {et~al.} 2013, The Astrophysical Journal, 769, 3, \dodoi{10.1088/0004-637x/769/1/3}

\bibitem[{Narayanan {et~al.}(2020)Narayanan, Turk, Robitaille, Kelly, McClellan, Sharma, Garg, Abruzzo, Choi, Conroy, {et~al.}}]{powderday}
Narayanan, D., Turk, M.~J., Robitaille, T., {et~al.} 2020, arXiv preprint arXiv:2006.10757

\bibitem[{Oey {et~al.}(2007)Oey, Meurer, Yelda, Furst, Caballero‐Nieves, Hanish, Levesque, Thilker, Walth, Bland‐Hawthorn, Dopita, Ferguson, Heckman, Doyle, Drinkwater, Freeman, Kennicutt, Kilborn, Knezek, Koribalski, Meyer, Putman, Ryan‐Weber, Smith, Staveley‐Smith, Webster, Werk, \& Zwaan}]{oey_2007}
Oey, M.~S., Meurer, G.~R., Yelda, S., {et~al.} 2007, The Astrophysical Journal, 661, 801, \dodoi{10.1086/517867}

\bibitem[{Onodera {et~al.}(2016)Onodera, Carollo, Lilly, Renzini, Arimoto, Capak, Daddi, Scoville, Tacchella, Tatehora, \& Zamorani}]{onodera2016}
Onodera, M., Carollo, C.~M., Lilly, S., {et~al.} 2016, The Astrophysical Journal, 822, 42, \dodoi{10.3847/0004-637X/822/1/42}

\bibitem[{{Osterbrock} \& {Ferland}(2006)}]{osterbrook2006}
{Osterbrock}, D.~E., \& {Ferland}, G.~J. 2006, {Astrophysics of gaseous nebulae and active galactic nuclei}

\bibitem[{Pagel {et~al.}(1992)Pagel, Simonson, Terlevich, \& Edmunds}]{Pagel1992}
Pagel, B.~E.~J., Simonson, E.~A., Terlevich, R.~J., \& Edmunds, M.~G. 1992, \mnras, 255, 325, \dodoi{10.1093/mnras/255.2.325}

\bibitem[{Patr\'icio {et~al.}(2018)Patr\'icio, Christensen, Rhodin, Cañameras, \& Lara-López}]{Patricio2018}
Patr\'icio, V., Christensen, L., Rhodin, H., Cañameras, R., \& Lara-López, M.~A. 2018, Monthly Notices of the Royal Astronomical Society, 481, 3520, \dodoi{10.1093/mnras/sty2508}

\bibitem[{Perez-Montero(2014)}]{perez-montero_2014}
Perez-Montero, E. 2014, Monthly Notices of the Royal Astronomical Society, 441, 2663, \dodoi{10.1093/mnras/stu753}

\bibitem[{P{\'{e}}rez-Montero(2017)}]{PrezMontero_2017}
P{\'{e}}rez-Montero, E. 2017, Publications of the Astronomical Society of the Pacific, 129, 043001, \dodoi{10.1088/1538-3873/aa5abb}

\bibitem[{Pettini \& Pagel(2004)}]{Pettini}
Pettini, M., \& Pagel, B. E.~J. 2004, Monthly Notices of the Royal Astronomical Society, 348, L59

\bibitem[{Pilyugin {et~al.}(2012)Pilyugin, V{\'\i}lchez, Mattsson, \& Thuan}]{pilyugin2012}
Pilyugin, L., V{\'\i}lchez, J., Mattsson, L., \& Thuan, T. 2012, Monthly Notices of the Royal Astronomical Society, 421, 1624

\bibitem[{{Pilyugin, L. S.}(2005)}]{Pilyugin2005}
{Pilyugin, L. S.} 2005, A\&A, 436, L1, \dodoi{10.1051/0004-6361:200500108}

\bibitem[{Péroux \& Howk(2020)}]{peroux_cosmic_2020}
Péroux, C., \& Howk, J.~C. 2020, Annual Review of Astronomy and Astrophysics, 58, 363, \dodoi{10.1146/annurev-astro-021820-120014}

\bibitem[{Rahmati {et~al.}(2013)Rahmati, Pawlik, Raicevic, \& Schaye}]{rahmati_2013}
Rahmati, A., Pawlik, A.~H., Raicevic, M., \& Schaye, J. 2013, Monthly Notices of the Royal Astronomical Society, 430, 2427, \dodoi{10.1093/mnras/stt066}

\bibitem[{Reddy {et~al.}(2023{\natexlab{a}})Reddy, Topping, Sanders, Shapley, \& Brammer}]{reddy_2023a}
Reddy, N.~A., Topping, M.~W., Sanders, R.~L., Shapley, A.~E., \& Brammer, G. 2023{\natexlab{a}}, The Astrophysical Journal, 952, 167, \dodoi{10.3847/1538-4357/acd754}

\bibitem[{Reddy {et~al.}(2023{\natexlab{b}})Reddy, Sanders, Shapley, Topping, Kriek, Coil, Mobasher, Siana, \& Rezaee}]{reddy_2023b}
Reddy, N.~A., Sanders, R.~L., Shapley, A.~E., {et~al.} 2023{\natexlab{b}}, The Astrophysical Journal, 951, 56, \dodoi{10.3847/1538-4357/acd0b1}

\bibitem[{Rhoads {et~al.}(2023)Rhoads, Wold, Harish, Kim, Pharo, Malhotra, Gabrielpillai, Jiang, \& Yang}]{rhoads_2023}
Rhoads, J.~E., Wold, I. G.~B., Harish, S., {et~al.} 2023, The Astrophysical Journal Letters, 942, L14, \dodoi{10.3847/2041-8213/acaaaf}

\bibitem[{Rigby {et~al.}(2011)Rigby, Wuyts, Gladders, Sharon, \& Becker}]{rigby2011}
Rigby, J.~R., Wuyts, E., Gladders, M., Sharon, K., \& Becker, G. 2011, The Astrophysical Journal, 732, 59

\bibitem[{{Robitaille}(2011)}]{Hyperion}
{Robitaille}, T.~P. 2011, \aap, 536, A79, \dodoi{10.1051/0004-6361/201117150}

\bibitem[{Runco {et~al.}(2021)Runco, Shapley, Sanders, Topping, Kriek, Reddy, Coil, Mobasher, Siana, Freeman, Shivaei, Azadi, Price, Leung, Fetherolf, de~Groot, Zick, Fornasini, \& Barro}]{Runco2021}
Runco, J.~N., Shapley, A.~E., Sanders, R.~L., {et~al.} 2021, Monthly Notices of the Royal Astronomical Society, 502, 2600, \dodoi{10.1093/mnras/stab119}

\bibitem[{Sanders {et~al.}(2023{\natexlab{a}})Sanders, Shapley, Topping, Reddy, \& Brammer}]{sanders_direct_2023}
Sanders, R.~L., Shapley, A.~E., Topping, M.~W., Reddy, N.~A., \& Brammer, G.~B. 2023{\natexlab{a}}, \dodoi{10.48550/arxiv.2303.08149}

\bibitem[{Sanders {et~al.}(2023{\natexlab{b}})Sanders, Shapley, Topping, Reddy, \& Brammer}]{sanders_2023}
---. 2023{\natexlab{b}}, \dodoi{10.48550/arxiv.2301.06696}

\bibitem[{Sanders {et~al.}(2017)Sanders, Shapley, Zhang, \& Yan}]{sanders2017biases}
Sanders, R.~L., Shapley, A.~E., Zhang, K., \& Yan, R. 2017, The Astrophysical Journal, 850, 136, \dodoi{10.3847/1538-4357/aa93e4}

\bibitem[{Sanders {et~al.}(2016)Sanders, Shapley, Kriek, Reddy, Freeman, Coil, Siana, Mobasher, Shivaei, Price, \& de~Groot}]{Sanders_2016}
Sanders, R.~L., Shapley, A.~E., Kriek, M., {et~al.} 2016, The Astrophysical Journal, 825, L23, \dodoi{10.3847/2041-8205/825/2/l23}

\bibitem[{Sanders {et~al.}(2018)Sanders, Shapley, Kriek, Freeman, Reddy, Siana, Coil, Mobasher, Davé, Shivaei, Azadi, Price, Leung, Fetherholf, Groot, Zick, Fornasini, \& Barro}]{sanders_mosdef_2018}
---. 2018, The Astrophysical Journal, 858, 99, \dodoi{10.3847/1538-4357/aabcbd}

\bibitem[{Sanders {et~al.}(2019)Sanders, Shapley, Reddy, Kriek, Siana, Coil, Mobasher, Shivaei, Freeman, Azadi, Price, Leung, Fetherolf, de Groot, Zick, Fornasini, \& Barro}]{Sanders_direct2019}
Sanders, R.~L., Shapley, A.~E., Reddy, N.~A., {et~al.} 2019, Monthly Notices of the Royal Astronomical Society, 491, 1427, \dodoi{10.1093/mnras/stz3032}

\bibitem[{Schaerer {et~al.}(2022)Schaerer, Marques-Chaves, Barrufet, Oesch, Izotov, Naidu, Guseva, \& Brammer}]{schaerer_2022}
Schaerer, D., Marques-Chaves, R., Barrufet, L., {et~al.} 2022, Astronomy \& Astrophysics, 665, L4, \dodoi{10.1051/0004-6361/202244556}

\bibitem[{Shapley {et~al.}(2023{\natexlab{a}})Shapley, Reddy, Sanders, Topping, \& Brammer}]{shapley_2023a}
Shapley, A.~E., Reddy, N.~A., Sanders, R.~L., Topping, M.~W., \& Brammer, G.~B. 2023{\natexlab{a}}, The Astrophysical Journal Letters, 950, L1, \dodoi{10.3847/2041-8213/acd939}

\bibitem[{Shapley {et~al.}(2023{\natexlab{b}})Shapley, Sanders, Reddy, Topping, \& Brammer}]{shapley_2023b}
Shapley, A.~E., Sanders, R.~L., Reddy, N.~A., Topping, M.~W., \& Brammer, G.~B. 2023{\natexlab{b}}, The Astrophysical Journal, 954, 157, \dodoi{10.3847/1538-4357/acea5a}

\bibitem[{Shapley {et~al.}(2019)Shapley, Sanders, Shao, Reddy, Kriek, Coil, Mobasher, Siana, Shivaei, Freeman, Azadi, Price, Leung, Fetherolf, de~Groot, Zick, Fornasini, \& Barro}]{Shapley_2019}
Shapley, A.~E., Sanders, R.~L., Shao, P., {et~al.} 2019, The Astrophysical Journal, 881, L35, \dodoi{10.3847/2041-8213/ab385a}

\bibitem[{Shirazi {et~al.}(2014)Shirazi, Brinchmann, \& Rahmati}]{shirazi2014}
Shirazi, M., Brinchmann, J., \& Rahmati, A. 2014, The Astrophysical Journal, 787, 120, \dodoi{10.1088/0004-637X/787/2/120}

\bibitem[{Smith {et~al.}(2017)Smith, Bryan, Glover, Goldbaum, Turk, Regan, Wise, Schive, Abel, Emerick, {et~al.}}]{smith2017grackle}
Smith, B.~D., Bryan, G.~L., Glover, S.~C., {et~al.} 2017, Monthly Notices of the Royal Astronomical Society, 466, 2217

\bibitem[{Stanghellini {et~al.}(2009)Stanghellini, Lee, Shaw, Balick, \& Villaver}]{stanghellini_2009}
Stanghellini, L., Lee, T.-H., Shaw, R.~A., Balick, B., \& Villaver, E. 2009, The Astrophysical Journal, 702, 733, \dodoi{10.1088/0004-637X/702/1/733}

\bibitem[{Stanghellini {et~al.}(2005)Stanghellini, Shaw, \& Gilmore}]{stanghellini_2005}
Stanghellini, L., Shaw, R.~A., \& Gilmore, D. 2005, The Astrophysical Journal, 622, 294, \dodoi{10.1086/427912}

\bibitem[{Stanway \& Eldridge(2018)}]{stanway2018}
Stanway, E., \& Eldridge, J. 2018, Monthly Notices of the Royal Astronomical Society, 479, 75

\bibitem[{Stark {et~al.}(2014)Stark, Richard, Siana, Charlot, Freeman, Gutkin, Wofford, Robertson, Amanullah, Watson, \& Milvang-Jensen}]{Stark2014}
Stark, D.~P., Richard, J., Siana, B., {et~al.} 2014, Monthly Notices of the Royal Astronomical Society, 445, 3200, \dodoi{10.1093/mnras/stu1618}

\bibitem[{Stasińska {et~al.}(1998)Stasińska, Richer, \& McCall}]{stasinska_1998}
Stasińska, G., Richer, M.~G., \& McCall, M.~L. 1998, Astronomy and Astrophysics, 336, 667.
\newblock \url{https://ui.adsabs.harvard.edu/abs/1998A&A...336..667S}

\bibitem[{Steidel {et~al.}(2016)Steidel, Strom, Pettini, Rudie, Reddy, \& Trainor}]{steidel2016}
Steidel, C.~C., Strom, A.~L., Pettini, M., {et~al.} 2016, The Astrophysical Journal, 826, 159

\bibitem[{Steidel {et~al.}(2014)Steidel, Rudie, Strom, Pettini, Reddy, Shapley, Trainor, Erb, Turner, Konidaris, {et~al.}}]{steidel2014}
Steidel, C.~C., Rudie, G.~C., Strom, A.~L., {et~al.} 2014, The Astrophysical Journal, 795, 165

\bibitem[{Strom {et~al.}(2018)Strom, Steidel, Rudie, Trainor, \& Pettini}]{strom2018}
Strom, A.~L., Steidel, C.~C., Rudie, G.~C., Trainor, R.~F., \& Pettini, M. 2018, The Astrophysical Journal, 868, 117

\bibitem[{Sun {et~al.}(2022)Sun, Egami, Pirzkal, Rieke, Baum, Boyer, Boyett, Bunker, Cameron, Curti, Eisenstein, Gennaro, Greene, Jaffe, Kelly, Koekemoer, Kumari, Maiolino, Maseda, Perna, Rest, Robertson, Schlawin, Smit, Stansberry, Sunnquist, Tacchella, Williams, \& Willmer}]{sun_2022}
Sun, F., Egami, E., Pirzkal, N., {et~al.} 2022, First {Sample} of {H}\${\textbackslash}alpha\$+[{O} {III}] \${\textbackslash}lambda\$5007 {Line} {Emitters} at \$z {\textgreater} 6\$ through {JWST}/{NIRCam} {Slitless} {Spectroscopy}: {Physical} {Properties} and {Line} {Luminosity} {Functions},  arXiv, \dodoi{10.48550/arXiv.2209.03374}

\bibitem[{Tacchella {et~al.}(2022)Tacchella, Johnson, Robertson, Carniani, D'Eugenio, Kumar, Maiolino, Nelson, Suess, Übler, Williams, Adebusola, Alberts, Arribas, Bhatawdekar, Bonaventura, Bowler, Bunker, Cameron, Curti, Egami, Eisenstein, Frye, Hainline, Helton, Ji, Looser, Lyu, Perna, Rawle, Rieke, Rieke, Saxena, Sandles, Shivaei, Simmonds, Sun, Willmer, Willott, \& Witstok}]{tacchella_2022}
Tacchella, S., Johnson, B.~D., Robertson, B.~E., {et~al.} 2022, {JWST} {NIRCam}+{NIRSpec}: {Interstellar} medium and stellar populations of young galaxies with rising star formation and evolving gas reservoirs,  arXiv, \dodoi{10.48550/arXiv.2208.03281}

\bibitem[{Tang {et~al.}(2023)Tang, Stark, Chen, Mason, Topping, Endsley, Senchyna, Plat, Lu, Whitler, Robertson, \& Charlot}]{tang_2023}
Tang, M., Stark, D.~P., Chen, Z., {et~al.} 2023, {JWST}/{NIRSpec} {Spectroscopy} of \$z=7-9\$ {Star} {Forming} {Galaxies} with {CEERS}: {New} {Insight} into {Bright} {Ly}\${\textbackslash}alpha\$ {Emitters} in {Ionized} {Bubbles},  arXiv, \dodoi{10.48550/arXiv.2301.07072}

\bibitem[{Taylor {et~al.}(2022)Taylor, Barger, \& Cowie}]{taylor_2022}
Taylor, A.~J., Barger, A.~J., \& Cowie, L.~L. 2022, The Astrophysical Journal Letters, 939, L3, \dodoi{10.3847/2041-8213/ac959d}

\bibitem[{Thompson(2014)}]{Thompson2014}
Thompson, R. 2014, {pyGadgetReader: GADGET snapshot reader for python}.
\newblock \doeprint{1411.001}

\bibitem[{Topping {et~al.}(2020{\natexlab{a}})Topping, Shapley, Reddy, Sanders, Coil, Kriek, Mobasher, \& Siana}]{Topping2020a}
Topping, M.~W., Shapley, A.~E., Reddy, N.~A., {et~al.} 2020{\natexlab{a}}, arXiv preprint, arXiv:2008.02282, \dodoi{https://arxiv.org/abs/2008.02282}

\bibitem[{Topping {et~al.}(2020{\natexlab{b}})Topping, Shapley, Reddy, Sanders, Coil, Kriek, Mobasher, \& Siana}]{Topping2020b}
---. 2020{\natexlab{b}}, Monthly Notices of the Royal Astronomical Society, 495, 4430, \dodoi{10.1093/mnras/staa1410}

\bibitem[{Tremonti {et~al.}(2004)Tremonti, Heckman, Kauffmann, Brinchmann, Charlot, White, Seibert, Peng, Schlegel, Uomoto, {et~al.}}]{tremonti2004}
Tremonti, C.~A., Heckman, T.~M., Kauffmann, G., {et~al.} 2004, The Astrophysical Journal, 613, 898

\bibitem[{Trump {et~al.}(2022)Trump, Haro, Simons, Backhaus, Amorín, Dickinson, Fernández, Papovich, Nicholls, Kewley, Brunker, Salzer, Wilkins, Almaini, Bagley, Berg, Bhatawdekar, Bisigello, Buat, Burgarella, Calabrò, Casey, Ciesla, Cleri, Cole, Cooper, Cooray, Costantin, Ferguson, Finkelstein, Fujimoto, Gardner, Gawiser, Giavalisco, Grazian, Grogin, Hathi, Hirschmann, Holwerda, Huertas-Company, Hutchison, Jogee, Juneau, Jung, Kartaltepe, Kirkpatrick, Koekemoer, Lotz, Lucas, Magnelli, Matharu, Pérez-González, Pirzkal, Rafelski, Rose, Seillé, Somerville, Straughn, Tacchella, Vanderhoof, Weiner, Wuyts, Yung, \& Zavala}]{trump_2022}
Trump, J.~R., Haro, P.~A., Simons, R.~C., {et~al.} 2022, The {Physical} {Conditions} of {Emission}-{Line} {Galaxies} at {Cosmic} {Dawn} from {JWST}/{NIRSpec} {Spectroscopy} in the {SMACS} 0723 {Early} {Release} {Observations},  arXiv, \dodoi{10.48550/arXiv.2207.12388}

\bibitem[{Vale~Asari {et~al.}(2019)Vale~Asari, Couto, Cid~Fernandes, Stasi{\'n}ska, de~Amorim, Ruschel-Dutra, Werle, \& Florido}]{vale2019}
Vale~Asari, N., Couto, G., Cid~Fernandes, R., {et~al.} 2019, Monthly Notices of the Royal Astronomical Society, 489, 4721

\bibitem[{Villar-Martín {et~al.}(2004)Villar-Martín, Cerviño, \& González~Delgado}]{villar2004}
Villar-Martín, M., Cerviño, M., \& González~Delgado, R.~M. 2004, Monthly Notices of the Royal Astronomical Society, 355, 1132, \dodoi{10.1111/j.1365-2966.2004.08395.x}

\bibitem[{Wang {et~al.}(2022)Wang, Jones, Vulcani, Treu, Morishita, Roberts-Borsani, Malkan, Henry, Brammer, Strait, Bradač, Boyett, Calabrò, Castellano, Fontana, Glazebrook, Kelly, Leethochawalit, Marchesini, Santini, Trenti, \& Yang}]{wang_2022}
Wang, X., Jones, T., Vulcani, B., {et~al.} 2022, The Astrophysical Journal Letters, 938, L16, \dodoi{10.3847/2041-8213/ac959e}

\bibitem[{Westmoquette {et~al.}(2013)Westmoquette, James, Monreal-Ibero, \& Walsh}]{Westmoquette2013}
Westmoquette, M.~S., James, B., Monreal-Ibero, A., \& Walsh, J.~R. 2013, A\&A, 550, A88, \dodoi{10.1051/0004-6361/201220580}

\bibitem[{Williams {et~al.}(2023)Williams, Kelly, Chen, Brammer, Zitrin, Treu, Scarlata, Koekemoer, Oguri, Lin, Diego, Nonino, Hjorth, Langeroodi, Broadhurst, Rogers, Perez-Fournon, Foley, Jha, Filippenko, Strolger, Pierel, Poidevin, \& Yang}]{williams_2023}
Williams, H., Kelly, P.~L., Chen, W., {et~al.} 2023, A {Highly} {Magnified} and {Extremely} {Compact} {Galaxy} at {Redshift} 9.51 with {Strong} {Nebular} {Emission},  arXiv, \dodoi{10.48550/arXiv.2210.15699}

\bibitem[{Wuyts {et~al.}(2012)Wuyts, Rigby, Gladders, Gilbank, Sharon, Gralla, \& Bayliss}]{wuyts2012}
Wuyts, E., Rigby, J.~R., Gladders, M.~D., {et~al.} 2012, The Astrophysical Journal, 745, 86, \dodoi{10.1088/0004-637x/745/1/86}

\bibitem[{Yuan \& Kewley(2009)}]{Yuan_2009}
Yuan, T.-T., \& Kewley, L.~J. 2009, The Astrophysical Journal, 699, L161, \dodoi{10.1088/0004-637x/699/2/l161}

\bibitem[{Zaritsky {et~al.}(1994)Zaritsky, Kennicutt~Jr, \& Huchra}]{zaritsky1994}
Zaritsky, D., Kennicutt~Jr, R.~C., \& Huchra, J.~P. 1994, The Astrophysical Journal, 420, 87

\end{thebibliography}
\appendix
\section{Theoretical strong line calibrations using our model at high redshifts}\label{app:mi_higZ}
\begin{longtblr}[
  caption = {Theoretical strong line calibrations for z = 1 to 5 using the base model and for z = 5 using the model with a higher ionization parameter for different line ratios. We fit polynomials of the form $y = a_2*x^2 + a_1*x^1 + a_0$ where x is the metallicity (12 + log(O/H)) and y is the line ratio.},
  label = {tab:mi_highz},
]
{
  colspec = {|c|c|c|c|c|c|},
  rowhead = 1,
  row{2-7} = {m, cyan9},
  row{8-13} = {m},
  row{14-19} = {m, cyan9},
  row{20-25} = {m},
  row{26-31} = {m, cyan9},
  row{32-37} = {m},
  row{38-43} = {m, cyan9},
  row{44-49} = {m},
  row{50-55} = {m, cyan9},
  row{56-61} = {m},
  row{62-67} = {m, cyan9},
  row{68-73} = {m},
  row{1} = {olive9}
}
\hline
\textbf{Line Ratio} & \textbf{Redshift} & \textbf{Fit} & \textbf{Lower CI} & \textbf{Upper CI} & \textbf{Metallicity Range} \\
\hline
\SetCell[r=6]{}{\textbf{N2}} & 1 &{$a_0$: -6.662 \\ $a_1$: 0.138 \\ $a_2$: 0.061} & {$a_0$: -6.233 \\ $a_1$: 0.105 \\ $a_2$: 0.063} & {$a_0$: -7.091 \\ $a_1$: 0.17 \\ $a_2$: 0.059} & 7.3 - 8.9 \\ \cline{1-6} 
& 2 & {$a_0$: 16.247 \\ $a_1$: -5.606 \\ $a_2$: 0.42} & {$a_0$: 16.596 \\ $a_1$: -5.632 \\ $a_2$: 0.421} & {$a_0$: 15.899 \\ $a_1$: -5.581 \\ $a_2$: 0.418} & 7.1 - 8.7 \\ \cline{1-6} 
& 3 & {$a_0$: 9.368 \\ $a_1$: -3.923 \\ $a_2$: 0.316} & {$a_0$: 9.682 \\ $a_1$: -3.942 \\ $a_2$: 0.317} & {$a_0$: 9.055 \\ $a_1$: -3.904 \\ $a_2$: 0.315} & 7.0 - 8.6 \\ \cline{1-6} 
& 4 & {$a_0$: 10.87 \\ $a_1$: -4.264 \\ $a_2$: 0.335} & {$a_0$: 11.181 \\ $a_1$: -4.281 \\ $a_2$: 0.336} & {$a_0$: 10.559 \\ $a_1$: -4.248 \\ $a_2$: 0.334} & 6.9 - 8.5 \\ \cline{1-6} 
& 5 & {$a_0$: 6.902 \\ $a_1$: -3.176 \\ $a_2$: 0.261} & {$a_0$: 7.212 \\ $a_1$: -3.195 \\ $a_2$: 0.262} & {$a_0$: 6.593 \\ $a_1$: -3.156 \\ $a_2$: 0.26} & 6.8 - 8.4 \\ \cline{1-6} 
& 5 (high logU) & {$a_0$: 17.348 \\ $a_1$: -5.817 \\ $a_2$: 0.423} & {$a_0$: 17.933 \\ $a_1$: -5.9 \\ $a_2$: 0.429} & {$a_0$: 16.762 \\ $a_1$: -5.733 \\ $a_2$: 0.418} & 6.8 - 8.4 \\ \cline{1-6} 
\hline
\SetCell[r=5]{}{\textbf{R3}} & 1 &{$a_0$: -52.137 \\ $a_1$: 13.162 \\ $a_2$: -0.821} & {$a_0$: -51.754 \\ $a_1$: 13.133 \\ $a_2$: -0.819} & {$a_0$: -52.521 \\ $a_1$: 13.192 \\ $a_2$: -0.823} & 7.3 - 8.9 \\ \cline{1-6} 
& 2 & {$a_0$: -41.669 \\ $a_1$: 10.511 \\ $a_2$: -0.653} & {$a_0$: -41.425 \\ $a_1$: 10.494 \\ $a_2$: -0.652} & {$a_0$: -41.913 \\ $a_1$: 10.529 \\ $a_2$: -0.654} & 7.1 - 8.7 \\ \cline{1-6} 
& 3 & {$a_0$: -35.338 \\ $a_1$: 8.88 \\ $a_2$: -0.548} & {$a_0$: -35.114 \\ $a_1$: 8.867 \\ $a_2$: -0.547} & {$a_0$: -35.562 \\ $a_1$: 8.894 \\ $a_2$: -0.549} & 7.0 - 8.6 \\ \cline{1-6} 
& 4 & {$a_0$: -30.177 \\ $a_1$: 7.546 \\ $a_2$: -0.462} & {$a_0$: -29.971 \\ $a_1$: 7.535 \\ $a_2$: -0.461} & {$a_0$: -30.384 \\ $a_1$: 7.557 \\ $a_2$: -0.463} & 6.9 - 8.5 \\ \cline{1-6} 
& 5 & {$a_0$: -24.688 \\ $a_1$: 6.118 \\ $a_2$: -0.369} & {$a_0$: -24.473 \\ $a_1$: 6.104 \\ $a_2$: -0.368} & {$a_0$: -24.902 \\ $a_1$: 6.132 \\ $a_2$: -0.37} & 6.8 - 8.4 \\ \cline{1-6} 
\textbf{R3} & 5 (high logU) & {$a_0$: -26.244 \\ $a_1$: 6.585 \\ $a_2$: -0.401} & {$a_0$: -25.914 \\ $a_1$: 6.538 \\ $a_2$: -0.399} & {$a_0$: -26.573 \\ $a_1$: 6.632 \\ $a_2$: -0.404} & 6.8 - 8.4 \\ \cline{1-6} 
\hline
\SetCell[r=6]{}{\textbf{R2}} & 1 &{$a_0$: -35.496 \\ $a_1$: 8.42 \\ $a_2$: -0.497} & {$a_0$: -35.222 \\ $a_1$: 8.399 \\ $a_2$: -0.495} & {$a_0$: -35.771 \\ $a_1$: 8.441 \\ $a_2$: -0.498} & 7.3 - 8.9 \\ \cline{1-6} 
& 2 & {$a_0$: -21.727 \\ $a_1$: 4.973 \\ $a_2$: -0.281} & {$a_0$: -21.466 \\ $a_1$: 4.954 \\ $a_2$: -0.28} & {$a_0$: -21.989 \\ $a_1$: 4.992 \\ $a_2$: -0.283} & 7.1 - 8.7 \\ \cline{1-6} 
& 3 & {$a_0$: -24.442 \\ $a_1$: 5.618 \\ $a_2$: -0.32} & {$a_0$: -24.188 \\ $a_1$: 5.603 \\ $a_2$: -0.319} & {$a_0$: -24.696 \\ $a_1$: 5.634 \\ $a_2$: -0.321} & 7.0 - 8.6 \\ \cline{1-6} 
& 4 & {$a_0$: -19.516 \\ $a_1$: 4.391 \\ $a_2$: -0.244} & {$a_0$: -19.254 \\ $a_1$: 4.377 \\ $a_2$: -0.243} & {$a_0$: -19.777 \\ $a_1$: 4.405 \\ $a_2$: -0.245} & 6.9 - 8.5 \\ \cline{1-6} 
& 5 & {$a_0$: -19.95 \\ $a_1$: 4.556 \\ $a_2$: -0.258} & {$a_0$: -19.667 \\ $a_1$: 4.538 \\ $a_2$: -0.257} & {$a_0$: -20.233 \\ $a_1$: 4.574 \\ $a_2$: -0.259} & 6.8 - 8.4 \\ \cline{1-6} 
& 5 (high logU) & {$a_0$: -11.204 \\ $a_1$: 2.311 \\ $a_2$: -0.119} & {$a_0$: -10.668 \\ $a_1$: 2.235 \\ $a_2$: -0.114} & {$a_0$: -11.74 \\ $a_1$: 2.388 \\ $a_2$: -0.124} & 6.8 - 8.4 \\ \cline{1-6} 
\hline
\SetCell[r=6]{}{\textbf{R23}} & 1 &{$a_0$: -42.197 \\ $a_1$: 10.625 \\ $a_2$: -0.656} & {$a_0$: -41.955 \\ $a_1$: 10.606 \\ $a_2$: -0.655} & {$a_0$: -42.439 \\ $a_1$: 10.643 \\ $a_2$: -0.657} & 7.3 - 8.9 \\ \cline{1-6} 
& 2 & {$a_0$: -34.364 \\ $a_1$: 8.641 \\ $a_2$: -0.53} & {$a_0$: -34.18 \\ $a_1$: 8.628 \\ $a_2$: -0.529} & {$a_0$: -34.548 \\ $a_1$: 8.655 \\ $a_2$: -0.531} & 7.1 - 8.7 \\ \cline{1-6} 
& 3 & {$a_0$: -31.514 \\ $a_1$: 7.9 \\ $a_2$: -0.482} & {$a_0$: -31.324 \\ $a_1$: 7.889 \\ $a_2$: -0.481} & {$a_0$: -31.704 \\ $a_1$: 7.911 \\ $a_2$: -0.483} & 7.0 - 8.6 \\ \cline{1-6} 
& 4 & {$a_0$: -27.149 \\ $a_1$: 6.777 \\ $a_2$: -0.41} & {$a_0$: -26.973 \\ $a_1$: 6.768 \\ $a_2$: -0.41} & {$a_0$: -27.325 \\ $a_1$: 6.787 \\ $a_2$: -0.411} & 6.9 - 8.5 \\ \cline{1-6} 
& 5 & {$a_0$: -23.269 \\ $a_1$: 5.776 \\ $a_2$: -0.346} & {$a_0$: -23.077 \\ $a_1$: 5.764 \\ $a_2$: -0.345} & {$a_0$: -23.46 \\ $a_1$: 5.788 \\ $a_2$: -0.346} & 6.8 - 8.4 \\ \cline{1-6} 
& 5 (high logU) & {$a_0$: -24.758 \\ $a_1$: 6.233 \\ $a_2$: -0.378} & {$a_0$: -24.446 \\ $a_1$: 6.189 \\ $a_2$: -0.376} & {$a_0$: -25.07 \\ $a_1$: 6.278 \\ $a_2$: -0.381} & 6.8 - 8.4 \\ \cline{1-6} 
\hline
\SetCell[r=6]{}{\textbf{O3N2}} & 1 &{$a_0$: -45.45 \\ $a_1$: 13.017 \\ $a_2$: -0.881} & {$a_0$: -44.783 \\ $a_1$: 12.966 \\ $a_2$: -0.878} & {$a_0$: -46.116 \\ $a_1$: 13.068 \\ $a_2$: -0.885} & 7.3 - 8.9 \\ \cline{1-6} 
& 2 & {$a_0$: -57.392 \\ $a_1$: 15.982 \\ $a_2$: -1.064} & {$a_0$: -56.912 \\ $a_1$: 15.947 \\ $a_2$: -1.062} & {$a_0$: -57.871 \\ $a_1$: 16.016 \\ $a_2$: -1.066} & 7.1 - 8.7 \\ \cline{1-6} 
& 3 & {$a_0$: -44.706 \\ $a_1$: 12.803 \\ $a_2$: -0.864} & {$a_0$: -44.303 \\ $a_1$: 12.779 \\ $a_2$: -0.863} & {$a_0$: -45.11 \\ $a_1$: 12.827 \\ $a_2$: -0.866} & 7.0 - 8.6 \\ \cline{1-6} 
& 4 & {$a_0$: -41.047 \\ $a_1$: 11.811 \\ $a_2$: -0.797} & {$a_0$: -40.658 \\ $a_1$: 11.789 \\ $a_2$: -0.796} & {$a_0$: -41.436 \\ $a_1$: 11.832 \\ $a_2$: -0.798} & 6.9 - 8.5 \\ \cline{1-6} 
& 5 & {$a_0$: -31.59 \\ $a_1$: 9.294 \\ $a_2$: -0.63} & {$a_0$: -31.208 \\ $a_1$: 9.27 \\ $a_2$: -0.629} & {$a_0$: -31.972 \\ $a_1$: 9.318 \\ $a_2$: -0.632} & 6.8 - 8.4 \\ \cline{1-6} 
& 5 (high logU) & {$a_0$: -43.591 \\ $a_1$: 12.402 \\ $a_2$: -0.825} & {$a_0$: -42.887 \\ $a_1$: 12.302 \\ $a_2$: -0.818} & {$a_0$: -44.295 \\ $a_1$: 12.502 \\ $a_2$: -0.831} & 6.8 - 8.4 \\ \cline{1-6} 
\hline
\SetCell[r=6]{}{\textbf{R3N2}} & 1 &{$a_0$: -48.618 \\ $a_1$: 13.809 \\ $a_2$: -0.939} & {$a_0$: -47.876 \\ $a_1$: 13.752 \\ $a_2$: -0.935} & {$a_0$: -49.359 \\ $a_1$: 13.866 \\ $a_2$: -0.942} & 7.3 - 8.9 \\ \cline{1-6} 
& 2 & {$a_0$: -64.199 \\ $a_1$: 17.678 \\ $a_2$: -1.177} & {$a_0$: -63.674 \\ $a_1$: 17.64 \\ $a_2$: -1.175} & {$a_0$: -64.724 \\ $a_1$: 17.716 \\ $a_2$: -1.18} & 7.1 - 8.7 \\ \cline{1-6} 
& 3 & {$a_0$: -50.615 \\ $a_1$: 14.258 \\ $a_2$: -0.961} & {$a_0$: -50.181 \\ $a_1$: 14.232 \\ $a_2$: -0.959} & {$a_0$: -51.049 \\ $a_1$: 14.284 \\ $a_2$: -0.963} & 7.0 - 8.6 \\ \cline{1-6} 
& 4 & {$a_0$: -45.803 \\ $a_1$: 12.962 \\ $a_2$: -0.874} & {$a_0$: -45.39 \\ $a_1$: 12.94 \\ $a_2$: -0.873} & {$a_0$: -46.215 \\ $a_1$: 12.984 \\ $a_2$: -0.875} & 6.9 - 8.5 \\ \cline{1-6} 
& 5 & {$a_0$: -35.253 \\ $a_1$: 10.157 \\ $a_2$: -0.688} & {$a_0$: -34.853 \\ $a_1$: 10.132 \\ $a_2$: -0.686} & {$a_0$: -35.653 \\ $a_1$: 10.182 \\ $a_2$: -0.69} & 6.8 - 8.4 \\ \cline{1-6} 
& 5 (high logU) & {$a_0$: -46.413 \\ $a_1$: 13.038 \\ $a_2$: -0.868} & {$a_0$: -45.683 \\ $a_1$: 12.934 \\ $a_2$: -0.861} & {$a_0$: -47.143 \\ $a_1$: 13.143 \\ $a_2$: -0.874} & 6.8 - 8.4 \\ \cline{1-6} 
\hline
\textbf{N2O2} & 1 &{$a_0$: 31.541 \\ $a_1$: -8.959 \\ $a_2$: 0.608} & {$a_0$: 32.024 \\ $a_1$: -8.996 \\ $a_2$: 0.61} & {$a_0$: 31.057 \\ $a_1$: -8.922 \\ $a_2$: 0.606} & 7.3 - 8.9 \\ \cline{1-6} 
\SetCell[r=5]{}{\textbf{N2O2}} & 2 & {$a_0$: 44.395 \\ $a_1$: -12.176 \\ $a_2$: 0.808} & {$a_0$: 44.698 \\ $a_1$: -12.198 \\ $a_2$: 0.809} & {$a_0$: 44.091 \\ $a_1$: -12.154 \\ $a_2$: 0.807} & 7.1 - 8.7 \\ \cline{1-6} 
& 3 & {$a_0$: 39.719 \\ $a_1$: -10.996 \\ $a_2$: 0.733} & {$a_0$: 39.964 \\ $a_1$: -11.011 \\ $a_2$: 0.734} & {$a_0$: 39.474 \\ $a_1$: -10.982 \\ $a_2$: 0.732} & 7.0 - 8.6 \\ \cline{1-6} 
& 4 & {$a_0$: 35.141 \\ $a_1$: -9.807 \\ $a_2$: 0.656} & {$a_0$: 35.37 \\ $a_1$: -9.819 \\ $a_2$: 0.657} & {$a_0$: 34.912 \\ $a_1$: -9.794 \\ $a_2$: 0.655} & 6.9 - 8.5 \\ \cline{1-6} 
& 5 & {$a_0$: 30.515 \\ $a_1$: -8.595 \\ $a_2$: 0.577} & {$a_0$: 30.767 \\ $a_1$: -8.611 \\ $a_2$: 0.578} & {$a_0$: 30.264 \\ $a_1$: -8.579 \\ $a_2$: 0.576} & 6.8 - 8.4 \\ \cline{1-6} 
& 5 (high logU) & {$a_0$: 31.373 \\ $a_1$: -8.764 \\ $a_2$: 0.585} & {$a_0$: 31.833 \\ $a_1$: -8.83 \\ $a_2$: 0.589} & {$a_0$: 30.913 \\ $a_1$: -8.699 \\ $a_2$: 0.581} & 6.8 - 8.4 \\ \cline{1-6} 
\hline
\SetCell[r=6]{}{\textbf{O3O2}} & 1 &{$a_0$: -16.862 \\ $a_1$: 4.797 \\ $a_2$: -0.328} & {$a_0$: -16.364 \\ $a_1$: 4.759 \\ $a_2$: -0.325} & {$a_0$: -17.36 \\ $a_1$: 4.835 \\ $a_2$: -0.33} & 7.3 - 8.9 \\ \cline{1-6} 
& 2 & {$a_0$: -19.924 \\ $a_1$: 5.534 \\ $a_2$: -0.371} & {$a_0$: -19.548 \\ $a_1$: 5.507 \\ $a_2$: -0.37} & {$a_0$: -20.3 \\ $a_1$: 5.561 \\ $a_2$: -0.373} & 7.1 - 8.7 \\ \cline{1-6} 
& 3 & {$a_0$: -10.896 \\ $a_1$: 3.262 \\ $a_2$: -0.228} & {$a_0$: -10.573 \\ $a_1$: 3.242 \\ $a_2$: -0.227} & {$a_0$: -11.219 \\ $a_1$: 3.281 \\ $a_2$: -0.229} & 7.0 - 8.6 \\ \cline{1-6} 
& 4 & {$a_0$: -10.661 \\ $a_1$: 3.155 \\ $a_2$: -0.218} & {$a_0$: -10.345 \\ $a_1$: 3.138 \\ $a_2$: -0.217} & {$a_0$: -10.978 \\ $a_1$: 3.172 \\ $a_2$: -0.219} & 6.9 - 8.5 \\ \cline{1-6} 
& 5 & {$a_0$: -4.737 \\ $a_1$: 1.562 \\ $a_2$: -0.111} & {$a_0$: -4.414 \\ $a_1$: 1.541 \\ $a_2$: -0.11} & {$a_0$: -5.061 \\ $a_1$: 1.582 \\ $a_2$: -0.113} & 6.8 - 8.4 \\ \cline{1-6} 
& 5 (high logU) & {$a_0$: -15.04 \\ $a_1$: 4.274 \\ $a_2$: -0.283} & {$a_0$: -14.453 \\ $a_1$: 4.19 \\ $a_2$: -0.277} & {$a_0$: -15.627 \\ $a_1$: 4.358 \\ $a_2$: -0.288} & 6.8 - 8.4 \\ \cline{1-6} 
\hline
\SetCell[r=2]{}{\textbf{Ne3O2}} & 1 &{$a_0$: -3.828 \\ $a_1$: 1.067 \\ $a_2$: -0.085} & {$a_0$: -3.489 \\ $a_1$: 1.04 \\ $a_2$: -0.084} & {$a_0$: -4.166 \\ $a_1$: 1.095 \\ $a_2$: -0.087} & 7.3 - 8.9 \\ \cline{1-6} 
& 2 & {$a_0$: -14.735 \\ $a_1$: 3.814 \\ $a_2$: -0.258} & {$a_0$: -14.416 \\ $a_1$: 3.79 \\ $a_2$: -0.256} & {$a_0$: -15.054 \\ $a_1$: 3.837 \\ $a_2$: -0.259} & 7.1 - 8.7 \\ \cline{1-6} 
\SetCell[r=4]{}{\textbf{Ne3O2}} & 3 & {$a_0$: -8.895 \\ $a_1$: 2.345 \\ $a_2$: -0.165} & {$a_0$: -8.611 \\ $a_1$: 2.328 \\ $a_2$: -0.164} & {$a_0$: -9.179 \\ $a_1$: 2.362 \\ $a_2$: -0.166} & 7.0 - 8.6 \\ \cline{1-6} 
& 4 & {$a_0$: -10.372 \\ $a_1$: 2.686 \\ $a_2$: -0.185} & {$a_0$: -10.084 \\ $a_1$: 2.671 \\ $a_2$: -0.184} & {$a_0$: -10.659 \\ $a_1$: 2.702 \\ $a_2$: -0.186} & 6.9 - 8.5 \\ \cline{1-6} 
& 5 & {$a_0$: -6.314 \\ $a_1$: 1.597 \\ $a_2$: -0.112} & {$a_0$: -6.011 \\ $a_1$: 1.578 \\ $a_2$: -0.11} & {$a_0$: -6.618 \\ $a_1$: 1.616 \\ $a_2$: -0.113} & 6.8 - 8.4 \\ \cline{1-6} 
& 5 (high logU) & {$a_0$: -14.295 \\ $a_1$: 3.768 \\ $a_2$: -0.252} & {$a_0$: -13.706 \\ $a_1$: 3.684 \\ $a_2$: -0.247} & {$a_0$: -14.884 \\ $a_1$: 3.852 \\ $a_2$: -0.257} & 6.8 - 8.4 \\ \cline{1-6} 
\hline
\SetCell[r=6]{}{\textbf{S2}} & 1 &{$a_0$: -20.957 \\ $a_1$: 4.132 \\ $a_2$: -0.207} & {$a_0$: -20.434 \\ $a_1$: 4.093 \\ $a_2$: -0.205} & {$a_0$: -21.48 \\ $a_1$: 4.172 \\ $a_2$: -0.209} & 7.3 - 8.9 \\ \cline{1-6} 
& 2 & {$a_0$: -1.276 \\ $a_1$: -0.78 \\ $a_2$: 0.098} & {$a_0$: -0.853 \\ $a_1$: -0.811 \\ $a_2$: 0.1} & {$a_0$: -1.699 \\ $a_1$: -0.75 \\ $a_2$: 0.096} & 7.1 - 8.7 \\ \cline{1-6} 
& 3 & {$a_0$: -8.202 \\ $a_1$: 0.94 \\ $a_2$: -0.01} & {$a_0$: -7.847 \\ $a_1$: 0.918 \\ $a_2$: -0.009} & {$a_0$: -8.557 \\ $a_1$: 0.961 \\ $a_2$: -0.012} & 7.0 - 8.6 \\ \cline{1-6} 
& 4 & {$a_0$: -6.348 \\ $a_1$: 0.532 \\ $a_2$: 0.011} & {$a_0$: -6.012 \\ $a_1$: 0.514 \\ $a_2$: 0.013} & {$a_0$: -6.685 \\ $a_1$: 0.55 \\ $a_2$: 0.01} & 6.9 - 8.5 \\ \cline{1-6} 
& 5 & {$a_0$: -10.901 \\ $a_1$: 1.8 \\ $a_2$: -0.076} & {$a_0$: -10.565 \\ $a_1$: 1.779 \\ $a_2$: -0.075} & {$a_0$: -11.236 \\ $a_1$: 1.821 \\ $a_2$: -0.078} & 6.8 - 8.4 \\ \cline{1-6} 
& 5 (high logU) & {$a_0$: -4.682 \\ $a_1$: 0.242 \\ $a_2$: 0.016} & {$a_0$: -4.134 \\ $a_1$: 0.164 \\ $a_2$: 0.021} & {$a_0$: -5.229 \\ $a_1$: 0.32 \\ $a_2$: 0.012} & 6.8 - 8.4 \\ \cline{1-6} 
\hline
\SetCell[r=3]{}{\textbf{O3S2}} & 1 &{$a_0$: -30.333 \\ $a_1$: 8.815 \\ $a_2$: -0.6} & {$a_0$: -29.541 \\ $a_1$: 8.754 \\ $a_2$: -0.596} & {$a_0$: -31.124 \\ $a_1$: 8.875 \\ $a_2$: -0.604} & 7.3 - 8.9 \\ \cline{1-6} 
& 2 & {$a_0$: -39.426 \\ $a_1$: 11.041 \\ $a_2$: -0.735} & {$a_0$: -38.874 \\ $a_1$: 11.001 \\ $a_2$: -0.732} & {$a_0$: -39.978 \\ $a_1$: 11.081 \\ $a_2$: -0.737} & 7.1 - 8.7 \\ \cline{1-6} 
& 3 & {$a_0$: -27.136 \\ $a_1$: 7.94 \\ $a_2$: -0.538} & {$a_0$: -26.695 \\ $a_1$: 7.914 \\ $a_2$: -0.536} & {$a_0$: -27.578 \\ $a_1$: 7.967 \\ $a_2$: -0.54} & 7.0 - 8.6 \\ \cline{1-6} 
\SetCell[r=3]{}{\textbf{O3S2}} & 4 & {$a_0$: -23.829 \\ $a_1$: 7.014 \\ $a_2$: -0.473} & {$a_0$: -23.419 \\ $a_1$: 6.992 \\ $a_2$: -0.472} & {$a_0$: -24.239 \\ $a_1$: 7.036 \\ $a_2$: -0.475} & 6.9 - 8.5 \\ \cline{1-6} 
& 5 & {$a_0$: -13.787 \\ $a_1$: 4.318 \\ $a_2$: -0.293} & {$a_0$: -13.392 \\ $a_1$: 4.293 \\ $a_2$: -0.292} & {$a_0$: -14.182 \\ $a_1$: 4.343 \\ $a_2$: -0.295} & 6.8 - 8.4 \\ \cline{1-6} 
& 5 (high logU) & {$a_0$: -21.562 \\ $a_1$: 6.343 \\ $a_2$: -0.418} & {$a_0$: -20.948 \\ $a_1$: 6.256 \\ $a_2$: -0.412} & {$a_0$: -22.176 \\ $a_1$: 6.431 \\ $a_2$: -0.423} & 6.8 - 8.4 \\ \cline{1-6} 
\hline
\SetCell[r=6]{}{\textbf{N2S2}} & 1 &{$a_0$: 12.618 \\ $a_1$: -3.979 \\ $a_2$: 0.286} & {$a_0$: 13.126 \\ $a_1$: -4.018 \\ $a_2$: 0.288} & {$a_0$: 12.11 \\ $a_1$: -3.941 \\ $a_2$: 0.283} & 7.3 - 8.9 \\ \cline{1-6} 
& 2 & {$a_0$: 21.596 \\ $a_1$: -6.25 \\ $a_2$: 0.429} & {$a_0$: 21.867 \\ $a_1$: -6.269 \\ $a_2$: 0.43} & {$a_0$: 21.325 \\ $a_1$: -6.23 \\ $a_2$: 0.428} & 7.1 - 8.7 \\ \cline{1-6} 
& 3 & {$a_0$: 20.043 \\ $a_1$: -5.898 \\ $a_2$: 0.41} & {$a_0$: 20.269 \\ $a_1$: -5.912 \\ $a_2$: 0.411} & {$a_0$: 19.818 \\ $a_1$: -5.885 \\ $a_2$: 0.409} & 7.0 - 8.6 \\ \cline{1-6} 
& 4 & {$a_0$: 20.088 \\ $a_1$: -5.922 \\ $a_2$: 0.412} & {$a_0$: 20.293 \\ $a_1$: -5.933 \\ $a_2$: 0.413} & {$a_0$: 19.883 \\ $a_1$: -5.911 \\ $a_2$: 0.411} & 6.9 - 8.5 \\ \cline{1-6} 
& 5 & {$a_0$: 19.625 \\ $a_1$: -5.814 \\ $a_2$: 0.406} & {$a_0$: 19.842 \\ $a_1$: -5.828 \\ $a_2$: 0.407} & {$a_0$: 19.409 \\ $a_1$: -5.801 \\ $a_2$: 0.405} & 6.8 - 8.4 \\ \cline{1-6} 
& 5 (high logU) & {$a_0$: 26.609 \\ $a_1$: -7.594 \\ $a_2$: 0.519} & {$a_0$: 27.039 \\ $a_1$: -7.656 \\ $a_2$: 0.522} & {$a_0$: 26.179 \\ $a_1$: -7.533 \\ $a_2$: 0.515} & 6.8 - 8.4 \\ \cline{1-6} 
\hline
\end{longtblr}
\end{document}